\documentclass[letterpaper,english,aps, prl, reprint, superscriptaddress]{revtex4-2}

\usepackage[T1]{fontenc}
\usepackage[latin9]{inputenc}
\usepackage{amsmath}
\usepackage{amssymb}
\usepackage{bbold}
\usepackage{graphicx}

\makeatletter

\pdfpageheight\paperheight
\pdfpagewidth\paperwidth

\PassOptionsToPackage{position=top}{subfig}
\usepackage{hyperref}
\hypersetup{
	colorlinks = true,
	allcolors = blue
}

\usepackage{times}

\makeatother

\usepackage{babel}
\begin{document}

\title{Exact multiple complex mobility edges and quantum state engineering in coupled 1D quasicystals}

\author{Li Wang}
\email{liwangiphy@sxu.edu.cn}
\affiliation{Institute of Theoretical Physics, State Key Laboratory of Quantum Optics Technologies and Devices, Collaborative Innovation Center of Extreme Optics, Shanxi University, Taiyuan 030006, China}

\author{Zhenbo Wang}
\affiliation{Institute of Theoretical Physics, State Key Laboratory of Quantum Optics Technologies and Devices, Collaborative Innovation Center of Extreme Optics, Shanxi University, Taiyuan 030006, China}
\affiliation{Institute of Advanced Functional Materials and Devices, Shanxi University, Taiyuan 030031, China}

\author{Jiaqi Liu}
\affiliation{Institute of Theoretical Physics, State Key Laboratory of Quantum Optics Technologies and Devices, Collaborative Innovation Center of Extreme Optics, Shanxi University, Taiyuan 030006, China}

\author{Shu Chen}
\affiliation{Beijing National Laboratory for Condensed Matter Physics, Institute
of Physics, Chinese Academy of Sciences, Beijing 100190, China}
\affiliation{School of Physical Sciences, University of Chinese Academy of Sciences,
Beijing 100049, China }

\date{\today}

\begin{abstract}
The key concept of mobility edge, which marks the critical transition between extended and localized states in energy domain,  has attracted significant interest in the cutting-edge frontiers of modern physics due to its profound implications for understanding localization and transport properties in disordered systems. 
However, a generic way to construct multiple mobility edges (MME) is still ambiguous and lacking.
In this work, we propose a brief scheme to engineer both real and complex exact multiple mobility edges exploiting a few coupled one-dimensional quasiperiodic chains. 
We study the extended-localized transitions of coupled one-dimensional quasiperiodic chains along the chain direction.
The model combines both the well-established quasiperiodicity and a kind of freshly introduced staggered non-reciprocity, which are aligned in two mutually perpendicular directions, within a unified framework.
Based on analytical analysis, we predict that 
when the couplings between quasiperiodic chains are weak,  the system will be in a mixed phase in which the localized states and extended states coexist and intertwine, thus lacking explicit energy separations. 
However, 
as  the inter-chain couplings increase to certain strength, exact multiple mobility edges emerge.
This prediction is clearly verified by concrete numerical calculations of the Fractal Dimension and the scaling index $\beta$.
Moreover, we show that the combination of quasiperiodicity and the staggered non-reciprocity can be utilized to design and realize quantum  states of various configurations.
Our results reveal a brief and general scheme to implement exact multiple mobility edges for synthetic materials engineering.
\end{abstract}

\maketitle

\textcolor{blue}{\em Introduction.}--
As is known, the mobility edge~\cite{Mott1967,Mott1987jpc,Abrahams1979prl,RevModPhys.57.287,RevModPhys.80.1355} is a pivotal concept in condensed matter physics, which represents the critical energy boundary between quantum states with distinct localization properties, such as localized states, extended states, and critical states. In-depth investigation of the mobility edge holds crucial significance for understanding the renowned Anderson localizations~\cite{Anderson1958pr},   transport properties~\cite{PhysRevLett.101.076803, sunQF}, and electronic structure of various materials.
Thus, enormous efforts from physicists has been made to study the fundamental mobility edge physics. Some of these studies focus on systems subject to uncorrelated random disorders~\cite{Mott1967,Mott1987jpc}, while others devote themselves to quasiperiodic lattices~\cite{AA1980,Harper_1955,Thouless1972,Thouless,Kohmoto,Kohmoto2008,Cai2013,Roati,Lahini,Bloch,wyc20prl,prb10817,LXJprl131.176401,wang2024MEprb,
li2024ringprb,flagellateprb}. 
Quasiperiodic lattices constitute a class of structures that is intermediate between periodic and truly random, which are nowadays being extensively investigated both theoretically and experimentially~\cite{PhysRevLett.116.140401,PhysRevX.7.041047,Bloch,An21prl,GAO2024}, particularly in one-dimensional (1D) and two-dimensional (2D) cases.  This is partially because, quasiperiodic lattices have been demonstrated to be able to  accommodate  extended-localized transitions even in 1D~\cite{AA1980,Harper_1955}, while in 1D and 2D truly random lattices, almost all states are localized and thus mobility edges  are absent~\cite{Abrahams1979prl,RevModPhys.57.287,RevModPhys.80.1355}. 

The introduction of the seminal Aubry-Andr\'{e}-Harper (AAH) model~\cite{AA1980,Harper_1955} has significantly spured the study of localization transitions and the fundamental mobility edge physics in low-dimensional systems. Numerous generalizations of the AAH model have been proposed, with some concentrating on short-range~\cite{Roy21prl,LXJprl131.176401}  or long-range hopping~\cite{Biddle10prl, Biddle11prb, Santos19prl, prb103075124}, while others are dedicated to designing intricate quasiperiodic potentials~\cite{xiexc1988prl,xieprb415544,Ganeshan2015prl,Lixp16prb,Sarma17prb,Lix20prb,GAO2024,prb105L220201,prb96174207,prb10817,wyc20prl}  to induce the emergence of mobility edges.
Moreover, the research on Anderson localization in the context of quasiperiodic lattices is no longer confined solely to the domain of Hermitian physics,  it has now expanded its boundaries into the booming field of non-Hermitian physics~\cite{bender98,RPP70947,NP1411,AP69249,NP131117,prl121086803,prl121.026808,JPC2035043,PRX8031079,
PRX9041015,PRB99201103,PRL123066404,NP16761,PRL125126402,
PRL124086801,CPB30020506,prl128120401,PRL124056802,PRL125226402,PRL127116801,NC115491,RMP93015005,PRX13021007,flagellateprb}.
The interplay of  parity-time ($\mathcal{PT}$) symmetry~\cite{bender98} and quasiperiodicity has been attracting intensive investigations~\cite{Chen21prb,PhysRevB.101.174205,PhysRevB.103.214202,Mishra,Longhi2019,HuHui,Gandhi,
Mishra,XueP,LiuYX2021a,Liuyx2021,Datta,ZhouLW,XiaX2022,PhysRevB.106.144208}. 
Notably, a highly favorable  observation is revealed that the real-complex transition in eigenenergy is closely connected to the  energy-dependent extended-localized transition in the system. Thus, the mobility edges for the  $\mathcal{PT}$-symmetric non-Hermitian  lattices are obtained in real energy domain.
On the other hand, the exploration of general non-Hermitian quasicrystals~\cite{wang2024MEprb,li2024ringprb,flagellateprb}, which are not restricted to specific symmetries, has naturally generalized the  concept to complex mobility edges by implementing the robust and exact Avila's global theory~\cite{avila,zhouqiwang2023} to the complex domain.
Recently, multiple mobility edges besides the traditional single mobility edge have begun to attract the attention of physicists. 
However, the topic is still in its infancy, with only a handful of studies~\cite{wyc20prl, GAO2024} exploring it. Up to now, there is still no systematic scheme for engineering multiple mobility edges in low dimensional systems, which leads to the conception of this work.

In this work, we investigate the localization properties of a lattice system composed of several coupled identical one-dimensional quasiperiodic chains. 
The coupling between the 1D quasiperiodic chains are staggered and non-reciprocal.
When the coupling between chains is relatively small, the coupled chain system can exist in extended phase, localized phase, or a intermediate phase~\cite{Sarma17prb}, depending on the strength of the quasiperiodic potential.
Interestingly, as the coupling between chains increases, clear and distinguishable multiple mobility edges separating localized states and extended states emerge. 
And analytical formula for the emergent multiple mobility edges can be exactly given as long as the original 1D quasiperiodic lattice possesses exact mobility edge.
Moreover,  the quasiperiodic potential in the 1D quasicrystal can be either Hermitian or non-Hermitian. 
For general non-Hermitian quasiperiodic potential  without explicit symmetry, multiple complex mobility edges are observed.
Thus, this provides a flexible  and generic  way to engineer exact multiple mobility edges by simply coupling identical 1D quasicrytals.
Furthermore, it is shown that various intriguing quantum states can be induced when several 1D quasiperiodic chains are coupled. 
By modulating the staggered non-reciprocal coupling, interesting quantum states like stripped states and point states emerge.

\begin{figure}[tp]
\includegraphics[width=8cm]{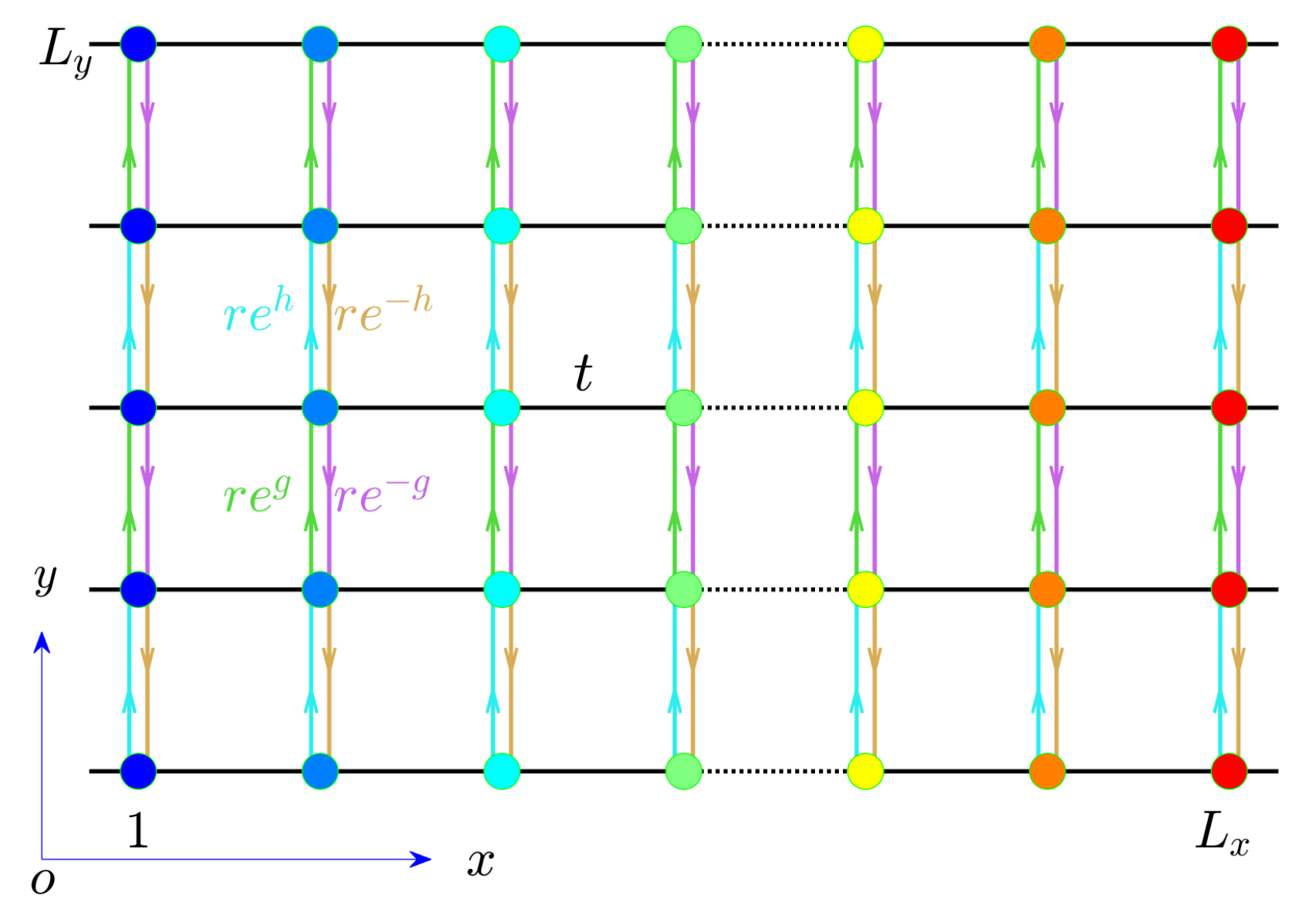}
\caption{Sketch of the coupled 1D quasicrystals.
Each small circle represents a lattice site. The variation in the color of the circles is used to indicate the difference in the on-site potential of each lattice site.
The vertical connecting lines in colors represent the staggered non-reciprocal hopping processes, while the horizontal black lines denote the normal hoppings $t$ along the $x$-direction.}
\label{Fig01}
\end{figure}

\textcolor{blue}{\em Model and exact multiple mobility edges.}--We study localization transitions in a class of coupled one-dimensional (1D) quasiperiodic lattices as depicted in Fig. \ref{Fig01}, which can be briefly described by the following tight-binding hamiltonian,
\begin{equation}
H=H_h+H_v, \label{C1DQc}
\end{equation}
with
\begin{align}
H_h&=\sum_{x}^{L_x}\sum_{y}^{L_y} \left[t \left(a^{\dagger}_{x,y} a_{x+1,y}+h.c.\right )+ V_x n_{x,y}\right],  \label{Hx}  \\
H_v&=\sum_{x}^{L_x} r \left [ \sum_{y \in odd}^{L_y-1} \left ( e^{-h} a^{\dagger}_{x,y} a_{x,y+1}+e^{h} a^{\dagger}_{x,y+1} a_{x,y} \right) \right.  \nonumber \\
&+\left. \sum_{y\in even }^{L_y-1} \left ( e^{-g} a^{\dagger}_{x,y} a_{x,y+1}+e^{g} a^{\dagger}_{x,y+1} a_{x,y} \right) \right ], \label{Hy}
\end{align}
where $a^{\dagger}_{x,y}$ and $a_{x,y}$ are the creation and annihilation operator of particles at site $(x,y)$, respectively, and $n_{x,y}$ is the corresponding particle number operator.
$t$ and $r$ denote the hopping amplitudes along $x$ and $y$ directions, respectively.
$t$ is set to be $1$ as the energy unit throughout this work.
The staggered nonreciprocal couplings between chains are described by parameters $h$ and $g$ , which respectively dictate the non-reciprocity of the hopping processes on the odd and even bonds in $y$ direction.
As will be seen later, they can be exploited to implement quantum state engineering conveniently.
$L_x$ and $L_y$ correspond to the dimensions of the coupled 1D quasicrystals in the $x$ and $y$ directions, respectively.
$V_x$ represents the one-dimensional on-site quasi-periodic lattice potential, whose specific form can be variously chosen and set according to actual needs.
Notably, $V_x$  can be either purely real or complex.
To illustrate the main results of the two cases concretely and systematically, we consider a brief and unified expression for the on-site potential $V_x$  as an example, which reads~\cite{wang2024MEprb},
\begin{equation}\label{Vx}
V_x=\frac{\lambda e^{i\chi}\cos(2\pi\alpha x+ \varphi)+\delta}{1-b \cos(2\pi\alpha x + \varphi)},  
\end{equation} 
where $\lambda$ dictates the strength of the quasiperiodic potential and $b\in (-1,1)$  is a deformation parameter. 
The parameter $\varphi$ denotes a phase shift and $\alpha$ is an irrational number responsible for the quasiperiodicity of the on site potential. To be concrete, we choose $\alpha=(\sqrt{5}-1)/2$ in this work. The obtained results are also valid for any other choice of the irrational number $\alpha$.
The parameter $\chi$ is a phase angle that integrates the Hermitian  case and the non-Hermitian case  within a single framework for unified description.
The on-site potential $V_x$ is purely real and it is the reentrance of the well known Ganeshan-Pixley-Das Sarma (GPD) quasiperiodic potential~\cite{Ganeshan2015prl} when $\chi=m \pi$,  which is expoited in the Hermitian case of 1D quasicrytals in this work. 
When $\chi \neq m \pi$, the on-site potential $V_x$ is generally complex, which is used by the non-Hermitian case.
Notably, the parameter $\delta$ can effectively diversify the energy spectrum of the system while keeping the mobility edge unchanged. 
As shown in Fig. \ref{Fig01}, the lattice model considered in this work has periodic boundary condition (PBC) in the $x$-direction and open boundary condition (OBC) in the $y$-direction. 

\begin{figure}[tbp]
\includegraphics[width=8.7cm]{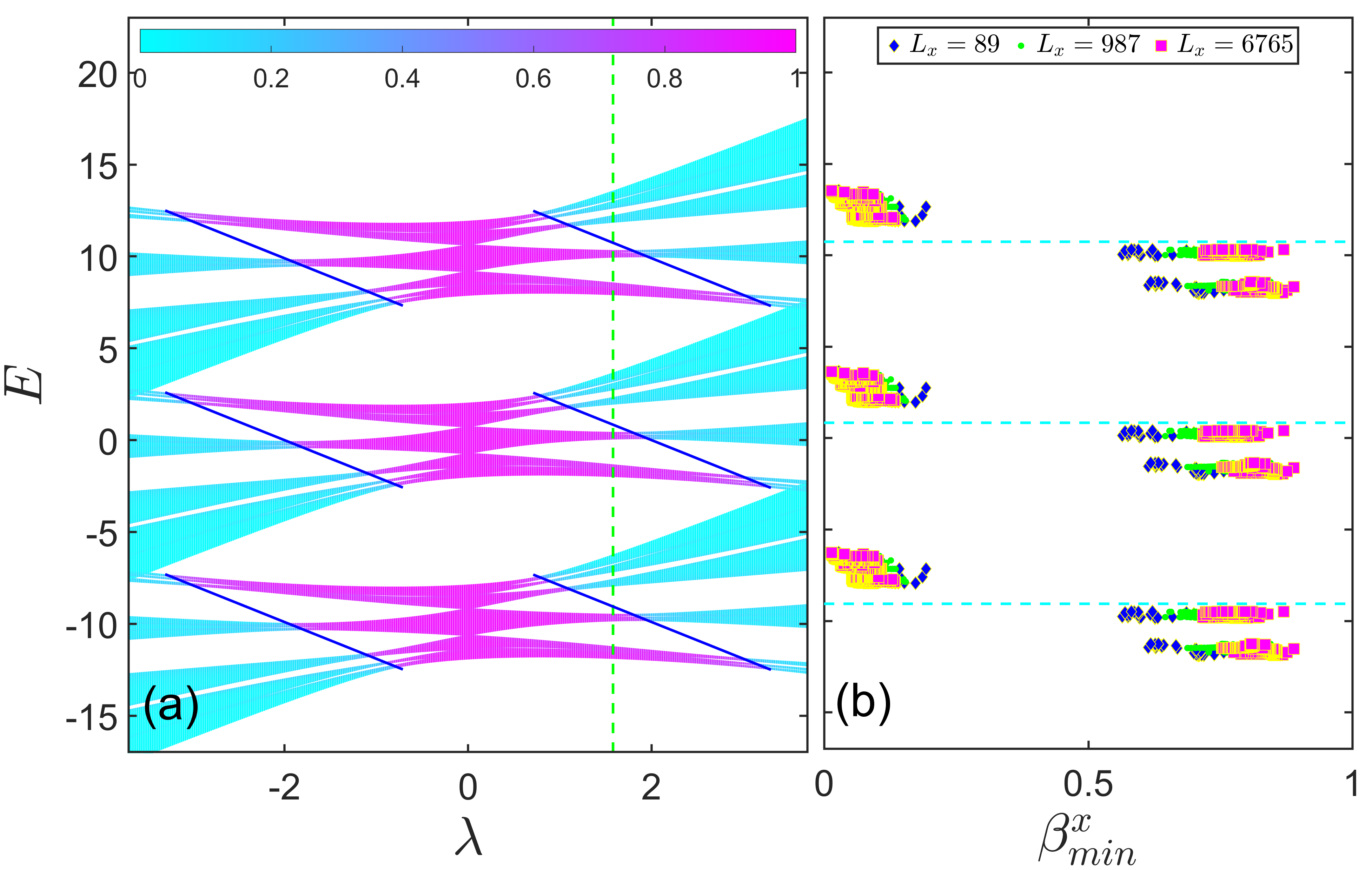}
\caption{\label{realMME}
(a) Representative energy spectra and exact real multiple mobility edges of the coupled 1D quasicrystals described by Eq.(\ref{C1DQc}) 
with $L_x=987$, $L_y=3$, $b=0.5$, $\chi=0$, $h=1$, $g=-1$, $r=7$, $\delta=0$, and $t=1$. 
The color of each energy point denotes the fraction dimension $F\!D_x$ of the corresponding eigenvector to visualize its localization property.
The three pairs of blue lines denote the exact real multiple mobility edges plotted according to the analytical formula Eq.~(\ref{mme}).
(b) The minimal scaling indices $\beta_{min}^x$s for all the eigenstates of the system at $\lambda=1.58$  which corresponds to the green dashed line in (a) .
Result with different lattice sizes $L_x=89$ and $L_x=6765$ denoted by filled diamonds and squares respectively  are also given to show tendency. 
The cyan dashed lines represent the multiple mobility edges at $\lambda=1.58$.
}\label{Fig02}
\end{figure}

Analytically, $H_v$ is amenable to exact diagonalization by a similar transformation $U$ through $U^{-1} H_v U$ with eigenvalue $\epsilon_y$. By virtue of this fact, the total Hamiltonian $H$  can be decoupled, yielding the following effective Hamiltonian,
\begin{align}
H_{eff}&=\sum_{y}^{L_y} \sum_{x}^{L_x} \left[t \left(a^{\dagger}_{x,y} a_{x+1,y}+h.c.\right )+ (V_x+\epsilon_y) n_{x,y}\right],  \label{Heff} 
\end{align}
with $\epsilon_y=2 r \cos(\frac{y \pi}{L_y+1})$.  From the viewpoint of $H_{eff}$, it is clear that $\epsilon_y$ acts as an energy shift to each $y$-chain with $y=1,2,...,L_y$ . It can be straightforwardly inferred that when the coupling strength $r$  is small, the energy spectra of different $y$-chains will overlap. 
When  $r$ is large enough, multiple complex mobility edges  will emerge, which can be described by an exact analytical formula. 
Specifically, for the concrete quasiperiodic potential $V_x$ in Eq. (\ref{Vx}), the analytical formula for the exact multiple complex mobility edges is as follows~\cite{wang2024MEprb},
\begin{align}
[b (E^r-\epsilon_y^r) + \lambda \cos \chi]^2 + \frac{[b (E^i-\epsilon_y^i) + \lambda \sin \chi]^2}{1 - b^2} = 4t^2, \label{mme}
\end{align}
where $E^r$($\epsilon_y^r$) and $E^i$($\epsilon_y^i$) are respectively the real and imaginary parts of the eigenenergy $E$($\epsilon_y$).
Numerical verification of this inference can be achieved by conducting specific calculations on lattices of finite sizes.

To carry out finite-size numerical analysis on the localization properties for the coupled 1D quasicrystals, we resort to  the fractal dimension (FD)~\cite{RevModPhys.80.1355, Hiramoto, PhysRevLett.84.3690, PhysRevLett.124.136405} as a convenient measure to distinguish between localized and extended eigenstates. As in the model of Eq. (\ref{C1DQc}), the quasiperiodicity exists in $x$-direction, we thus exploit the fractional dimension in $x$-direction $F\!D_x$ as a tool. For an arbitrary normalized eigenstate $\phi$ of the Hamiltonian in Eq.~(\ref{C1DQc}), $F\!D_x$ for a finite size system is defined as $F\!D_x=-\ln(\sum_{x} {P_x}^2)/\ln L_x$, where $P_{x}=\sum_y |\phi_{x,y}|^2$  with $\phi_{x,y}$ the probability amplitude on lattice site $(x,y)$. For a localized state, $F\!D_x\!\rightarrow\!0$, while for an extended state, $F\!D_x\!\rightarrow\!1$, as $L_x$ approaches $\infty$.

\begin{figure}[tbp]
\includegraphics[width=8.7cm]{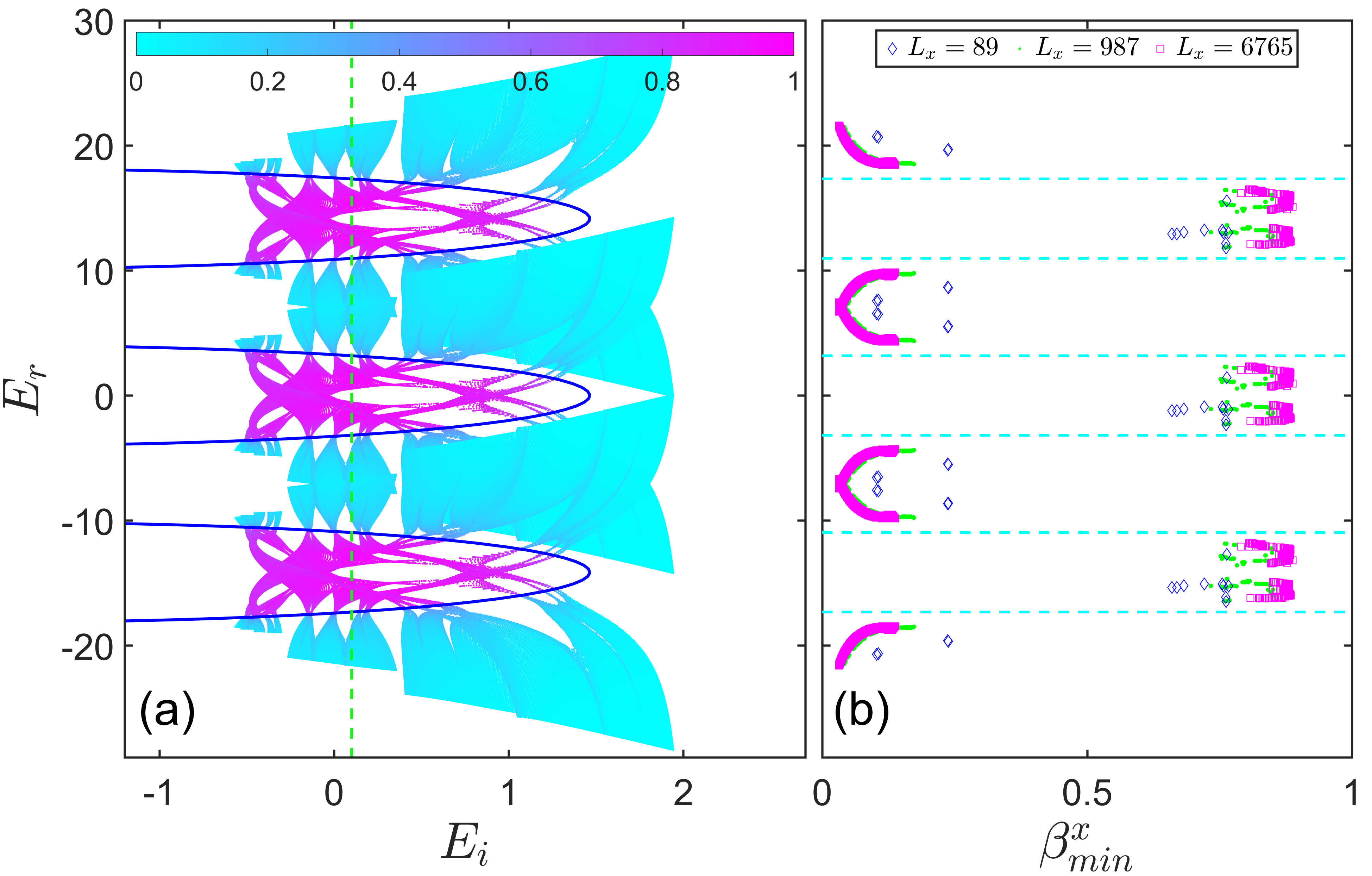}
\caption{\label{complexMME}
(a) Typical energy spectra and exact complex multiple mobility edges of the coupled 1D non-Hermitian quasicrystals described by Eq.(\ref{C1DQc}) with $L_x=987$, $L_y=3$, $b=0.5$, $\chi=0.5 \pi$, $h=1$, $g=1$, $r=10$, $\lambda=1$, and $t=1$. 
The color of each energy point denotes the fraction dimension $F\!D_x$ of the corresponding eigenvector to make its localization property visually perceivable.
The three blue ellipses represent the exact complex multiple mobility edges plotted according to the analytical formula Eq.~(\ref{mme}).
(b) The minimal scaling indices $\beta_{min}^x$s for all the eigenstates of the system with imaginary part of energy $E_i=0.1$  which corresponds to the green dashed line in (a) .
Result with different lattice sizes $L_x=89$ and $L_x=6765$ denoted by empty diamonds and squares respectively  are also given to show tendency. 
The cyan dashed lines represent the multiple mobility edges at $E_i=0.1$.
}\label{Fig03}
\end{figure}

Complementarily, we also perform finite-size calculations on the scaling index~\cite{RevModPhys.80.1355, Hiramoto, PhysRevB.93.104504,lin2023general} in $x$-direction $\beta^x$, which is defined through the probability distribution $P_{x}$ of a normalized wavefunction $\phi$, namely $P_{x}\propto L_x^{-\beta^{x}}$. 
It could be understood in an intuitive way as follows.
For an ideal extended eigenstate along the chain, $P_{x} \sim  1/L_x $ , thus one has $\beta^{x}=1$ for each postion $x$. For an ideal localized eigenstate, all the probability is concentrated on a single position $x$, while the probabilities on all other positions being zero. Therefore, the scaling index $\beta^x$ can only take two values,  namely $0$ and $\infty$.  Moreover, for a critical eigenstate, the minimal scaling index $\beta^x_{min}$ approaches to a value in the interval $(0,1)$.
In this way, the minimal scaling index $\beta^x_{min}$ can be used as a good criterion to distinguish between extended, critical, and  localized wavefunctions along $x$-direction.

\begin{figure*}[tbp]   
\includegraphics[width=18cm]{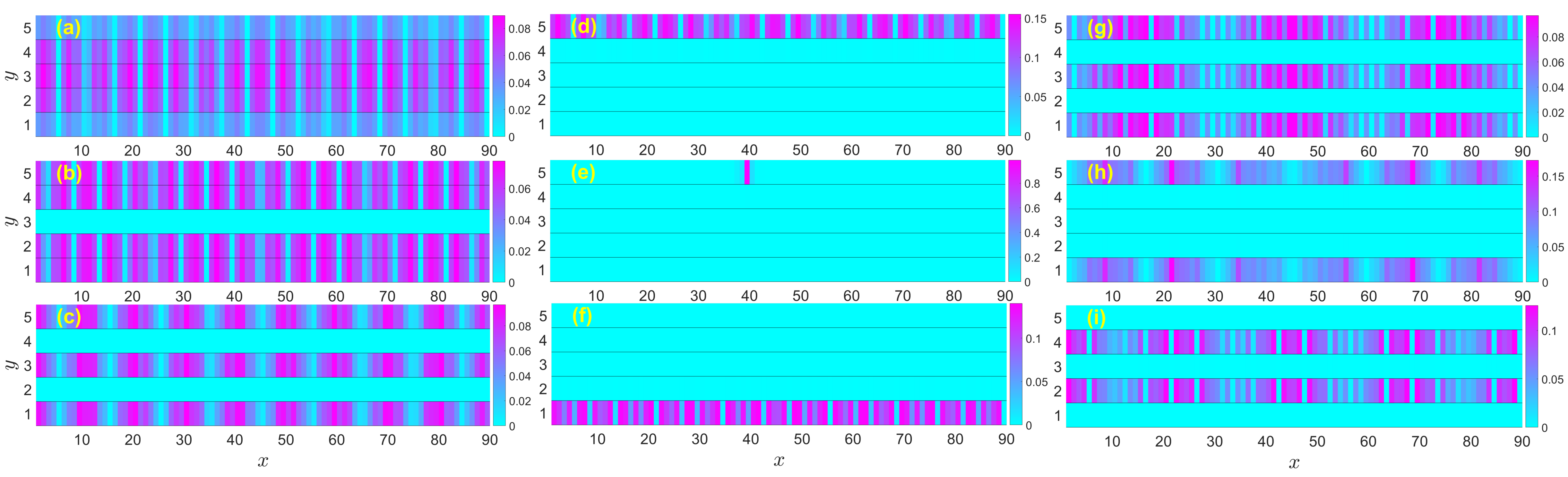}
\caption{\label{quantumstates}
Gallery of novel quantum states of the coupled 1D quasicrystals described by Eq.(\ref{C1DQc}) with $L_x=89$, $L_y=5$, $b=0.5$, $\chi=0$, $\lambda=1$, $r=20$, and $t=1$.
(a) The $17$th eigenstate with $h=g=0$,  $\delta=0$. 
(b) The $97$th eigenstate with $h=g=0$,  $\delta=0$. 
(c) The $187$th eigenstate with $h=g=0$,  $\delta=0$. 
(d) The $57$th eigenstate with $h=g=5$,  $\delta=0$. 
(e) The $205$th eigenstate with $h=g=5$,  $\delta=10$. 
(f) The $211$th eigenstate with $h=g=-5$,  $\delta=0$. 
(g) The $373$th eigenstate with $h=-g=5$,  $\delta=0$.
(h) The $187$th eigenstate with $h=-g=5$,  $\delta=0$.
(i) The $72$th eigenstate with $h=-g=-5$,  $\delta=0$.
The color in each sub-figure denotes the modulus of probability amplitude $|\phi_{x,y}|$ at each lattice site.
}\label{Fig04}
\end{figure*}

Firstly, we consider the Hermitian case in which the on-site quasiperiodic potential $V_x$ in each 1D quasicrystal is barely real with $\chi=m\pi$.  
We take coupled 1D quasicrytals with $L_x=987$ and $L_y=3$ as examples. In Fig. \ref{Fig02}a, we show the typcial phase diagram of the fractal dimension $F\!D_x$ in the ($E, \lambda$) plane with $b=0.5$, $\chi=0$, $h=1$, $g=-1$, $r=7$, $\delta=0$, and $t=1$. 
Clearly, reentrant localization transitions (RLT) in energy dimension show up. The series of the reentrant localization transition points are well described by the exact multiple mobility edges given by Eq. (\ref{mme}) which are denoted by three pairs of blue lines in Fig. \ref{Fig02}a.
Additionally, we also calculate the minimal scaling indices $\beta^x_{min}$s for all the eigenstates at $\lambda=1.58$ to visualize the reentrant localization transitions (RLT) in energy dimension. Apparently, taking the minmal scaling index $\beta^x_{min}$ as a criteria, one can find that extended states and localized states will appear alternatively as depicted in Fig. \ref{Fig02}b, which is in good agreement with the result shown in Fig. \ref{Fig02}a. To show the tendency of $\beta^x_{min}$ with lattice sizes, we also present the minimal scaling index for systems with $L_x=89$ and $L_x=6765$ for comparison.

Secondly, we explore the generically non-Hermitian case in which the parameter $\chi$  of the on-site quasiperiodic potential $V_x$  takes general values other than $m\pi$. As a concrete example, we numerically calculate the spectra for the non-Hermitian coupled 1D quasicrystals with $\delta$ varying from $-7$ to $7$. 
Other parameters are $L_x=987$, $L_y=3$, $b=0.5$, $\chi=0.5 \pi$, $h=1$, $g=1$, $r=10$, $\lambda=1$, and $t=1$.  
As depicted in Fig.~\ref{Fig03}a, the color of each point denotes the fractal dimension $F\!D_x$ of the corresponding eigenstate, which allows for an intuitive visualization of reentrant localization transitions (RTL) in the complex energy domain. 
Obviously,  the series of localized state regions and extended state regions can be well separated by the exact multiple complex mobility edges, which are denoted by three blue ellipses plotted according to the analytical formula Eq. (\ref{mme}).
Furthermore, in Fig. \ref{Fig03}b, we calculate the minimal scaling index $\beta^x_{min}$ for all states associated to eigenenergies with a fixed imaginary part $E_i$, denoted by a green dashed line in Fig. \ref{Fig03}a.
Evidently, with the real part of the eigenenergies $E_r$ increasing, the variation of the values of $\beta^x_{min}$ tells that the system enters localized phase and extended phase alternatively.  Regions with different localization properties are well separated by the multiple mobility edges, which are denoted by the cyan dashed lines. As expected, with the system size increasing from $L_x=89$ to $L_x=6765$, the minimal scaling index $\beta^x_{min}$ has the tendency to approach its ideal values 0 or 1.

\textcolor{blue}{\em Quantum state engineering via quasiperiodicity and the staggered non-reciprocity}
--The above-discussed scheme in this work not only provides a general approach to implement exact multiple mobility edges (MMEs), but also offers a brief route to achieve diverse and flexible control over quantum states.
Combining the quasiperiodic potentials in the one-dimensional quasicrystals and the staggered non-reciprocal coupling between them, one can conveniently realize many interesting and novel quantum states.  
In Fig. \ref{Fig04}, taking a finite lattice with $L_x=89$ and $L_y=5$ as a specific example, we demonstrate typical quantum states that may emerge in the system.
For the simplest case without non-reciprocity, i.e. $h=g=0$ and $\delta=0$, the lattice is in an extended phase.  It is shown that apart from the ordinary extended states like the one in Fig. \ref{Fig04}a, intriguing hollow states (Fig. \ref{Fig04}b) and sripped states (Fig. \ref{Fig04}c) have also made their appearance.
Interestingly, Fig. \ref{Fig04}b and Fig. \ref{Fig04}c are featured  by the existence of perfect empty quasiperiodic chains.  
Then if we turn on the non-reciprocity dictated by $h$ and $g$, skin effect~\cite{prl121086803} will come into play and all state will be pushed to the boundary, 
For extended states, we will obtain edge states, as shown in Fig. \ref{Fig04}d with $h=g=5$ and $\delta=0$. 
For localized states, one will get interesting point states, as depicted in Fig. \ref{Fig04}e with $h=g=5$ and $\delta=10$.
As expected, if we change the signs of  the non-reciprocity parameters $h$ and $g$, the accumulative direction of the probability will be reversed, as shown in Fig. \ref{Fig04}f with $h=g=-5$ and $\delta=0$.
Furthermore, we consider the most general case, i.e., the situation where the system has staggered non-reciprocal coupling in the $y$-direction. 
In this case, one can observe that the lattice exhibits forced stripped states, as shown in Fig. \ref{Fig04}g , and forced hollow states, as depicted in Fig. \ref{Fig04}h, in which $h=-g=5$ and $\delta=0$. 
Apparently, these states are termed forced states precisely because they are originated from the drive of the staggered non-reciprocal hoppings.
More interestingly, if we consider the reversed situation with $h=-g=-5$ and $\delta=0$, we obtain the complementary counterpart of the  striped state displayed in Fig. \ref{Fig04}g. It features empty edges, see Fig. \ref{Fig04}i. 
Other parameters in Fig. \ref{Fig04} are $b=0.5$, $\chi=0$, $\lambda=1$, $r=20$, and $t=1$.

\textcolor{blue}{\em Summary.}--
In this work, we have proposed a general and brief approach to realize multiple mobility edges (MMEs) in a highly controllable manner exploiting a few of 1D quasicrystals which are coupled by staggered non-reciprocal hoppings.
In our scheme, the quasiperiodic potential of 1D quasicrystals can be freely chosen. 
If it is purely real, one can obtain real exact multiple mobility edges, 
if it is complex, then one can get complex exact multiple mobility edges, as long as the original 1D quasicrytal possesses an exact mobility edge. 
Correspondingly, the analytical expression for the exact multiple mobility edges is explicitly given in the maintext.  
Furthermore, these findings are  verified by numerical computations of fractional dimension $F\!D_x$ and minimal scaling index $\beta_{min}^x$ .
Moreover, the staggered non-reciprocal coupling between 1D quasicrystals can also be used to realize flexible engineering on quantum states.
Intriguing quantum states like stripped states, hollow states and point states are revealed. 
This study will further facilitate the extension of the important concept of complex mobility edge to multiple case.

\textcolor{blue}{\em Acknowledgments}--
L.W. is supported by the Fundamental Research Program of Shanxi Province, China (Grant No. 202203021211315), Research Project Supported by Shanxi Scholarship Council of China (Grant No. 2024-004), the National Natural Science Foundation of China (Grant Nos. 11404199, 12147215) and the Fundamental Research Program of Shanxi Province, China (Grant Nos. 1331KSC and 2015021012). S. C. is supported by  by National Key Research and Development Program of China (Grant No. 2023YFA1406704), the NSFC under Grants No. 12174436 and
No. T2121001 and the Strategic Priority Research Program of Chinese Academy of Sciences under Grant No. XDB33000000.

%


\begin{thebibliography}{85}%
\makeatletter
\providecommand \@ifxundefined [1]{%
 \@ifx{#1\undefined}
}%
\providecommand \@ifnum [1]{%
 \ifnum #1\expandafter \@firstoftwo
 \else \expandafter \@secondoftwo
 \fi
}%
\providecommand \@ifx [1]{%
 \ifx #1\expandafter \@firstoftwo
 \else \expandafter \@secondoftwo
 \fi
}%
\providecommand \natexlab [1]{#1}%
\providecommand \enquote  [1]{``#1''}%
\providecommand \bibnamefont  [1]{#1}%
\providecommand \bibfnamefont [1]{#1}%
\providecommand \citenamefont [1]{#1}%
\providecommand \href@noop [0]{\@secondoftwo}%
\providecommand \href [0]{\begingroup \@sanitize@url \@href}%
\providecommand \@href[1]{\@@startlink{#1}\@@href}%
\providecommand \@@href[1]{\endgroup#1\@@endlink}%
\providecommand \@sanitize@url [0]{\catcode `\\12\catcode `\$12\catcode
  `\&12\catcode `\#12\catcode `\^12\catcode `\_12\catcode `\%12\relax}%
\providecommand \@@startlink[1]{}%
\providecommand \@@endlink[0]{}%
\providecommand \url  [0]{\begingroup\@sanitize@url \@url }%
\providecommand \@url [1]{\endgroup\@href {#1}{\urlprefix }}%
\providecommand \urlprefix  [0]{URL }%
\providecommand \Eprint [0]{\href }%
\providecommand \doibase [0]{https://doi.org/}%
\providecommand \selectlanguage [0]{\@gobble}%
\providecommand \bibinfo  [0]{\@secondoftwo}%
\providecommand \bibfield  [0]{\@secondoftwo}%
\providecommand \translation [1]{[#1]}%
\providecommand \BibitemOpen [0]{}%
\providecommand \bibitemStop [0]{}%
\providecommand \bibitemNoStop [0]{.\EOS\space}%
\providecommand \EOS [0]{\spacefactor3000\relax}%
\providecommand \BibitemShut  [1]{\csname bibitem#1\endcsname}%
\let\auto@bib@innerbib\@empty
\bibitem [{\citenamefont {Mott}(1967)}]{Mott1967}%
  \BibitemOpen
  \bibfield  {author} {\bibinfo {author} {\bibfnamefont {N.}~\bibnamefont
  {Mott}},\ }\bibfield  {title} {\bibinfo {title} {Electrons in disordered
  structures},\ }\href {https://doi.org/10.1080/00018736700101265} {\bibfield
  {journal} {\bibinfo  {journal} {Advances in Physics}\ }\textbf {\bibinfo
  {volume} {16}},\ \bibinfo {pages} {49} (\bibinfo {year} {1967})},\ \Eprint
  {https://arxiv.org/abs/https://doi.org/10.1080/00018736700101265}
  {https://doi.org/10.1080/00018736700101265} \BibitemShut {NoStop}%
\bibitem [{\citenamefont {Mott}(1987)}]{Mott1987jpc}%
  \BibitemOpen
  \bibfield  {author} {\bibinfo {author} {\bibfnamefont {N.}~\bibnamefont
  {Mott}},\ }\bibfield  {title} {\bibinfo {title} {The mobility edge since
  1967},\ }\href {https://doi.org/10.1088/0022-3719/20/21/008} {\bibfield
  {journal} {\bibinfo  {journal} {Journal of Physics C: Solid State Physics}\
  }\textbf {\bibinfo {volume} {20}},\ \bibinfo {pages} {3075} (\bibinfo {year}
  {1987})}\BibitemShut {NoStop}%
\bibitem [{\citenamefont {Abrahams}\ \emph {et~al.}(1979)\citenamefont
  {Abrahams}, \citenamefont {Anderson}, \citenamefont {Licciardello},\ and\
  \citenamefont {Ramakrishnan}}]{Abrahams1979prl}%
  \BibitemOpen
  \bibfield  {author} {\bibinfo {author} {\bibfnamefont {E.}~\bibnamefont
  {Abrahams}}, \bibinfo {author} {\bibfnamefont {P.~W.}\ \bibnamefont
  {Anderson}}, \bibinfo {author} {\bibfnamefont {D.~C.}\ \bibnamefont
  {Licciardello}},\ and\ \bibinfo {author} {\bibfnamefont {T.~V.}\ \bibnamefont
  {Ramakrishnan}},\ }\bibfield  {title} {\bibinfo {title} {Scaling theory of
  localization: Absence of quantum diffusion in two dimensions},\ }\href
  {https://doi.org/10.1103/PhysRevLett.42.673} {\bibfield  {journal} {\bibinfo
  {journal} {Phys. Rev. Lett.}\ }\textbf {\bibinfo {volume} {42}},\ \bibinfo
  {pages} {673} (\bibinfo {year} {1979})}\BibitemShut {NoStop}%
\bibitem [{\citenamefont {Lee}\ and\ \citenamefont
  {Ramakrishnan}(1985)}]{RevModPhys.57.287}%
  \BibitemOpen
  \bibfield  {author} {\bibinfo {author} {\bibfnamefont {P.~A.}\ \bibnamefont
  {Lee}}\ and\ \bibinfo {author} {\bibfnamefont {T.~V.}\ \bibnamefont
  {Ramakrishnan}},\ }\bibfield  {title} {\bibinfo {title} {Disordered
  electronic systems},\ }\href {https://doi.org/10.1103/RevModPhys.57.287}
  {\bibfield  {journal} {\bibinfo  {journal} {Rev. Mod. Phys.}\ }\textbf
  {\bibinfo {volume} {57}},\ \bibinfo {pages} {287} (\bibinfo {year}
  {1985})}\BibitemShut {NoStop}%
\bibitem [{\citenamefont {Evers}\ and\ \citenamefont
  {Mirlin}(2008)}]{RevModPhys.80.1355}%
  \BibitemOpen
  \bibfield  {author} {\bibinfo {author} {\bibfnamefont {F.}~\bibnamefont
  {Evers}}\ and\ \bibinfo {author} {\bibfnamefont {A.~D.}\ \bibnamefont
  {Mirlin}},\ }\bibfield  {title} {\bibinfo {title} {Anderson transitions},\
  }\href {https://doi.org/10.1103/RevModPhys.80.1355} {\bibfield  {journal}
  {\bibinfo  {journal} {Rev. Mod. Phys.}\ }\textbf {\bibinfo {volume} {80}},\
  \bibinfo {pages} {1355} (\bibinfo {year} {2008})}\BibitemShut {NoStop}%
\bibitem [{\citenamefont {Anderson}(1958)}]{Anderson1958pr}%
  \BibitemOpen
  \bibfield  {author} {\bibinfo {author} {\bibfnamefont {P.~W.}\ \bibnamefont
  {Anderson}},\ }\bibfield  {title} {\bibinfo {title} {Absence of diffusion in
  certain random lattices},\ }\href {https://doi.org/10.1103/PhysRev.109.1492}
  {\bibfield  {journal} {\bibinfo  {journal} {Phys. Rev.}\ }\textbf {\bibinfo
  {volume} {109}},\ \bibinfo {pages} {1492} (\bibinfo {year}
  {1958})}\BibitemShut {NoStop}%
\bibitem [{\citenamefont {Sil}\ \emph {et~al.}(2008)\citenamefont {Sil},
  \citenamefont {Maiti},\ and\ \citenamefont
  {Chakrabarti}}]{PhysRevLett.101.076803}%
  \BibitemOpen
  \bibfield  {author} {\bibinfo {author} {\bibfnamefont {S.}~\bibnamefont
  {Sil}}, \bibinfo {author} {\bibfnamefont {S.~K.}\ \bibnamefont {Maiti}},\
  and\ \bibinfo {author} {\bibfnamefont {A.}~\bibnamefont {Chakrabarti}},\
  }\bibfield  {title} {\bibinfo {title} {Metal-insulator transition in an
  aperiodic ladder network: An exact result},\ }\href
  {https://doi.org/10.1103/PhysRevLett.101.076803} {\bibfield  {journal}
  {\bibinfo  {journal} {Phys. Rev. Lett.}\ }\textbf {\bibinfo {volume} {101}},\
  \bibinfo {pages} {076803} (\bibinfo {year} {2008})}\BibitemShut {NoStop}%
\bibitem [{\citenamefont {Guo}\ \emph {et~al.}(2014)\citenamefont {Guo},
  \citenamefont {Xie},\ and\ \citenamefont {Sun}}]{sunQF}%
  \BibitemOpen
  \bibfield  {author} {\bibinfo {author} {\bibfnamefont {A.-M.}\ \bibnamefont
  {Guo}}, \bibinfo {author} {\bibfnamefont {X.~C.}\ \bibnamefont {Xie}},\ and\
  \bibinfo {author} {\bibfnamefont {Q.-f.}\ \bibnamefont {Sun}},\ }\bibfield
  {title} {\bibinfo {title} {Delocalization and scaling properties of
  low-dimensional quasiperiodic systems},\ }\href
  {https://doi.org/10.1103/PhysRevB.89.075434} {\bibfield  {journal} {\bibinfo
  {journal} {Phys. Rev. B}\ }\textbf {\bibinfo {volume} {89}},\ \bibinfo
  {pages} {075434} (\bibinfo {year} {2014})}\BibitemShut {NoStop}%
\bibitem [{\citenamefont {Aubry}\ and\ \citenamefont
  {Andr{\'e}}(1980)}]{AA1980}%
  \BibitemOpen
  \bibfield  {author} {\bibinfo {author} {\bibfnamefont {S.}~\bibnamefont
  {Aubry}}\ and\ \bibinfo {author} {\bibfnamefont {G.}~\bibnamefont
  {Andr{\'e}}},\ }\bibfield  {title} {\bibinfo {title} {Analyticity breaking
  and anderson localization in incommensurate lattices},\ }\href@noop {}
  {\bibfield  {journal} {\bibinfo  {journal} {Ann. Israel Phys. Soc}\ }\textbf
  {\bibinfo {volume} {3}},\ \bibinfo {pages} {18} (\bibinfo {year}
  {1980})}\BibitemShut {NoStop}%
\bibitem [{\citenamefont {Harper}(1955)}]{Harper_1955}%
  \BibitemOpen
  \bibfield  {author} {\bibinfo {author} {\bibfnamefont {P.~G.}\ \bibnamefont
  {Harper}},\ }\bibfield  {title} {\bibinfo {title} {Single band motion of
  conduction electrons in a uniform magnetic field},\ }\href
  {https://doi.org/10.1088/0370-1298/68/10/304} {\bibfield  {journal} {\bibinfo
   {journal} {Proceedings of the Physical Society. Section A}\ }\textbf
  {\bibinfo {volume} {68}},\ \bibinfo {pages} {874} (\bibinfo {year}
  {1955})}\BibitemShut {NoStop}%
\bibitem [{\citenamefont {Thouless}(1972)}]{Thouless1972}%
  \BibitemOpen
  \bibfield  {author} {\bibinfo {author} {\bibfnamefont {D.~J.}\ \bibnamefont
  {Thouless}},\ }\bibfield  {title} {\bibinfo {title} {A relation between the
  density of states and range of localization for one dimensional random
  systems},\ }\href {https://doi.org/10.1088/0022-3719/5/1/010} {\bibfield
  {journal} {\bibinfo  {journal} {Journal of Physics C: Solid State Physics}\
  }\textbf {\bibinfo {volume} {5}},\ \bibinfo {pages} {77} (\bibinfo {year}
  {1972})}\BibitemShut {NoStop}%
\bibitem [{\citenamefont {Thouless}(1988)}]{Thouless}%
  \BibitemOpen
  \bibfield  {author} {\bibinfo {author} {\bibfnamefont {D.~J.}\ \bibnamefont
  {Thouless}},\ }\bibfield  {title} {\bibinfo {title} {Localization by a
  potential with slowly varying period},\ }\href
  {https://doi.org/10.1103/PhysRevLett.61.2141} {\bibfield  {journal} {\bibinfo
   {journal} {Phys. Rev. Lett.}\ }\textbf {\bibinfo {volume} {61}},\ \bibinfo
  {pages} {2141} (\bibinfo {year} {1988})}\BibitemShut {NoStop}%
\bibitem [{\citenamefont {Kohmoto}(1983)}]{Kohmoto}%
  \BibitemOpen
  \bibfield  {author} {\bibinfo {author} {\bibfnamefont {M.}~\bibnamefont
  {Kohmoto}},\ }\bibfield  {title} {\bibinfo {title} {Metal-insulator
  transition and scaling for incommensurate systems},\ }\href
  {https://doi.org/10.1103/PhysRevLett.51.1198} {\bibfield  {journal} {\bibinfo
   {journal} {Phys. Rev. Lett.}\ }\textbf {\bibinfo {volume} {51}},\ \bibinfo
  {pages} {1198} (\bibinfo {year} {1983})}\BibitemShut {NoStop}%
\bibitem [{\citenamefont {Kohmoto}\ and\ \citenamefont
  {Tobe}(2008)}]{Kohmoto2008}%
  \BibitemOpen
  \bibfield  {author} {\bibinfo {author} {\bibfnamefont {M.}~\bibnamefont
  {Kohmoto}}\ and\ \bibinfo {author} {\bibfnamefont {D.}~\bibnamefont {Tobe}},\
  }\bibfield  {title} {\bibinfo {title} {Localization problem in a
  quasiperiodic system with spin-orbit interaction},\ }\href
  {https://doi.org/10.1103/PhysRevB.77.134204} {\bibfield  {journal} {\bibinfo
  {journal} {Phys. Rev. B}\ }\textbf {\bibinfo {volume} {77}},\ \bibinfo
  {pages} {134204} (\bibinfo {year} {2008})}\BibitemShut {NoStop}%
\bibitem [{\citenamefont {Cai}\ \emph {et~al.}(2013)\citenamefont {Cai},
  \citenamefont {Lang}, \citenamefont {Chen},\ and\ \citenamefont
  {Wang}}]{Cai2013}%
  \BibitemOpen
  \bibfield  {author} {\bibinfo {author} {\bibfnamefont {X.}~\bibnamefont
  {Cai}}, \bibinfo {author} {\bibfnamefont {L.-J.}\ \bibnamefont {Lang}},
  \bibinfo {author} {\bibfnamefont {S.}~\bibnamefont {Chen}},\ and\ \bibinfo
  {author} {\bibfnamefont {Y.}~\bibnamefont {Wang}},\ }\bibfield  {title}
  {\bibinfo {title} {Topological superconductor to anderson localization
  transition in one-dimensional incommensurate lattices},\ }\href
  {https://doi.org/10.1103/PhysRevLett.110.176403} {\bibfield  {journal}
  {\bibinfo  {journal} {Phys. Rev. Lett.}\ }\textbf {\bibinfo {volume} {110}},\
  \bibinfo {pages} {176403} (\bibinfo {year} {2013})}\BibitemShut {NoStop}%
\bibitem [{\citenamefont {Roati}\ \emph {et~al.}(2008)\citenamefont {Roati},
  \citenamefont {D'Errico}, \citenamefont {Fallani}, \citenamefont {Fattori},
  \citenamefont {Fort}, \citenamefont {Zaccanti}, \citenamefont {Modugno},
  \citenamefont {Modugno},\ and\ \citenamefont {Inguscio}}]{Roati}%
  \BibitemOpen
  \bibfield  {author} {\bibinfo {author} {\bibfnamefont {G.}~\bibnamefont
  {Roati}}, \bibinfo {author} {\bibfnamefont {C.}~\bibnamefont {D'Errico}},
  \bibinfo {author} {\bibfnamefont {L.}~\bibnamefont {Fallani}}, \bibinfo
  {author} {\bibfnamefont {M.}~\bibnamefont {Fattori}}, \bibinfo {author}
  {\bibfnamefont {C.}~\bibnamefont {Fort}}, \bibinfo {author} {\bibfnamefont
  {M.}~\bibnamefont {Zaccanti}}, \bibinfo {author} {\bibfnamefont
  {G.}~\bibnamefont {Modugno}}, \bibinfo {author} {\bibfnamefont
  {M.}~\bibnamefont {Modugno}},\ and\ \bibinfo {author} {\bibfnamefont
  {M.}~\bibnamefont {Inguscio}},\ }\bibfield  {title} {\bibinfo {title}
  {Anderson localization of a non-interacting bose-einstein condensate},\
  }\href {https://doi.org/10.1038/nature07071} {\bibfield  {journal} {\bibinfo
  {journal} {Nature}\ }\textbf {\bibinfo {volume} {453}},\ \bibinfo {pages}
  {895} (\bibinfo {year} {2008})}\BibitemShut {NoStop}%
\bibitem [{\citenamefont {Lahini}\ \emph {et~al.}(2009)\citenamefont {Lahini},
  \citenamefont {Pugatch}, \citenamefont {Pozzi}, \citenamefont {Sorel},
  \citenamefont {Morandotti}, \citenamefont {Davidson},\ and\ \citenamefont
  {Silberberg}}]{Lahini}%
  \BibitemOpen
  \bibfield  {author} {\bibinfo {author} {\bibfnamefont {Y.}~\bibnamefont
  {Lahini}}, \bibinfo {author} {\bibfnamefont {R.}~\bibnamefont {Pugatch}},
  \bibinfo {author} {\bibfnamefont {F.}~\bibnamefont {Pozzi}}, \bibinfo
  {author} {\bibfnamefont {M.}~\bibnamefont {Sorel}}, \bibinfo {author}
  {\bibfnamefont {R.}~\bibnamefont {Morandotti}}, \bibinfo {author}
  {\bibfnamefont {N.}~\bibnamefont {Davidson}},\ and\ \bibinfo {author}
  {\bibfnamefont {Y.}~\bibnamefont {Silberberg}},\ }\bibfield  {title}
  {\bibinfo {title} {Observation of a localization transition in quasiperiodic
  photonic lattices},\ }\href {https://doi.org/10.1103/PhysRevLett.103.013901}
  {\bibfield  {journal} {\bibinfo  {journal} {Phys. Rev. Lett.}\ }\textbf
  {\bibinfo {volume} {103}},\ \bibinfo {pages} {013901} (\bibinfo {year}
  {2009})}\BibitemShut {NoStop}%
\bibitem [{\citenamefont {L\"uschen}\ \emph {et~al.}(2018)\citenamefont
  {L\"uschen}, \citenamefont {Scherg}, \citenamefont {Kohlert}, \citenamefont
  {Schreiber}, \citenamefont {Bordia}, \citenamefont {Li}, \citenamefont
  {Das~Sarma},\ and\ \citenamefont {Bloch}}]{Bloch}%
  \BibitemOpen
  \bibfield  {author} {\bibinfo {author} {\bibfnamefont {H.~P.}\ \bibnamefont
  {L\"uschen}}, \bibinfo {author} {\bibfnamefont {S.}~\bibnamefont {Scherg}},
  \bibinfo {author} {\bibfnamefont {T.}~\bibnamefont {Kohlert}}, \bibinfo
  {author} {\bibfnamefont {M.}~\bibnamefont {Schreiber}}, \bibinfo {author}
  {\bibfnamefont {P.}~\bibnamefont {Bordia}}, \bibinfo {author} {\bibfnamefont
  {X.}~\bibnamefont {Li}}, \bibinfo {author} {\bibfnamefont {S.}~\bibnamefont
  {Das~Sarma}},\ and\ \bibinfo {author} {\bibfnamefont {I.}~\bibnamefont
  {Bloch}},\ }\bibfield  {title} {\bibinfo {title} {Single-particle mobility
  edge in a one-dimensional quasiperiodic optical lattice},\ }\href
  {https://doi.org/10.1103/PhysRevLett.120.160404} {\bibfield  {journal}
  {\bibinfo  {journal} {Phys. Rev. Lett.}\ }\textbf {\bibinfo {volume} {120}},\
  \bibinfo {pages} {160404} (\bibinfo {year} {2018})}\BibitemShut {NoStop}%
\bibitem [{\citenamefont {Wang}\ \emph {et~al.}(2020)\citenamefont {Wang},
  \citenamefont {Xia}, \citenamefont {Zhang}, \citenamefont {Yao},
  \citenamefont {Chen}, \citenamefont {You}, \citenamefont {Zhou},\ and\
  \citenamefont {Liu}}]{wyc20prl}%
  \BibitemOpen
  \bibfield  {author} {\bibinfo {author} {\bibfnamefont {Y.}~\bibnamefont
  {Wang}}, \bibinfo {author} {\bibfnamefont {X.}~\bibnamefont {Xia}}, \bibinfo
  {author} {\bibfnamefont {L.}~\bibnamefont {Zhang}}, \bibinfo {author}
  {\bibfnamefont {H.}~\bibnamefont {Yao}}, \bibinfo {author} {\bibfnamefont
  {S.}~\bibnamefont {Chen}}, \bibinfo {author} {\bibfnamefont {J.}~\bibnamefont
  {You}}, \bibinfo {author} {\bibfnamefont {Q.}~\bibnamefont {Zhou}},\ and\
  \bibinfo {author} {\bibfnamefont {X.-J.}\ \bibnamefont {Liu}},\ }\bibfield
  {title} {\bibinfo {title} {One-dimensional quasiperiodic mosaic lattice with
  exact mobility edges},\ }\href
  {https://doi.org/10.1103/PhysRevLett.125.196604} {\bibfield  {journal}
  {\bibinfo  {journal} {Phys. Rev. Lett.}\ }\textbf {\bibinfo {volume} {125}},\
  \bibinfo {pages} {196604} (\bibinfo {year} {2020})}\BibitemShut {NoStop}%
\bibitem [{\citenamefont {Wang}\ \emph
  {et~al.}(2023{\natexlab{a}})\citenamefont {Wang}, \citenamefont {Zhang},
  \citenamefont {Wang},\ and\ \citenamefont {Chen}}]{prb10817}%
  \BibitemOpen
  \bibfield  {author} {\bibinfo {author} {\bibfnamefont {Z.}~\bibnamefont
  {Wang}}, \bibinfo {author} {\bibfnamefont {Y.}~\bibnamefont {Zhang}},
  \bibinfo {author} {\bibfnamefont {L.}~\bibnamefont {Wang}},\ and\ \bibinfo
  {author} {\bibfnamefont {S.}~\bibnamefont {Chen}},\ }\bibfield  {title}
  {\bibinfo {title} {Engineering mobility in quasiperiodic lattices with exact
  mobility edges},\ }\href {https://doi.org/10.1103/PhysRevB.108.174202}
  {\bibfield  {journal} {\bibinfo  {journal} {Phys. Rev. B}\ }\textbf {\bibinfo
  {volume} {108}},\ \bibinfo {pages} {174202} (\bibinfo {year}
  {2023}{\natexlab{a}})}\BibitemShut {NoStop}%
\bibitem [{\citenamefont {Zhou}\ \emph {et~al.}(2023)\citenamefont {Zhou},
  \citenamefont {Wang}, \citenamefont {Poon}, \citenamefont {Zhou},\ and\
  \citenamefont {Liu}}]{LXJprl131.176401}%
  \BibitemOpen
  \bibfield  {author} {\bibinfo {author} {\bibfnamefont {X.-C.}\ \bibnamefont
  {Zhou}}, \bibinfo {author} {\bibfnamefont {Y.}~\bibnamefont {Wang}}, \bibinfo
  {author} {\bibfnamefont {T.-F.~J.}\ \bibnamefont {Poon}}, \bibinfo {author}
  {\bibfnamefont {Q.}~\bibnamefont {Zhou}},\ and\ \bibinfo {author}
  {\bibfnamefont {X.-J.}\ \bibnamefont {Liu}},\ }\bibfield  {title} {\bibinfo
  {title} {Exact new mobility edges between critical and localized states},\
  }\href {https://doi.org/10.1103/PhysRevLett.131.176401} {\bibfield  {journal}
  {\bibinfo  {journal} {Phys. Rev. Lett.}\ }\textbf {\bibinfo {volume} {131}},\
  \bibinfo {pages} {176401} (\bibinfo {year} {2023})}\BibitemShut {NoStop}%
\bibitem [{\citenamefont {Wang}\ \emph
  {et~al.}(2024{\natexlab{a}})\citenamefont {Wang}, \citenamefont {Wang},\ and\
  \citenamefont {Chen}}]{wang2024MEprb}%
  \BibitemOpen
  \bibfield  {author} {\bibinfo {author} {\bibfnamefont {L.}~\bibnamefont
  {Wang}}, \bibinfo {author} {\bibfnamefont {Z.}~\bibnamefont {Wang}},\ and\
  \bibinfo {author} {\bibfnamefont {S.}~\bibnamefont {Chen}},\ }\bibfield
  {title} {\bibinfo {title} {Non-hermitian butterfly spectra in a family of
  quasiperiodic lattices},\ }\href
  {https://doi.org/10.1103/PhysRevB.110.L060201} {\bibfield  {journal}
  {\bibinfo  {journal} {Phys. Rev. B}\ }\textbf {\bibinfo {volume} {110}},\
  \bibinfo {pages} {L060201} (\bibinfo {year}
  {2024}{\natexlab{a}})}\BibitemShut {NoStop}%
\bibitem [{\citenamefont {Li}\ and\ \citenamefont {Li}(2024)}]{li2024ringprb}%
  \BibitemOpen
  \bibfield  {author} {\bibinfo {author} {\bibfnamefont {S.-Z.}\ \bibnamefont
  {Li}}\ and\ \bibinfo {author} {\bibfnamefont {Z.}~\bibnamefont {Li}},\
  }\bibfield  {title} {\bibinfo {title} {Ring structure in the complex plane: A
  fingerprint of a non-hermitian mobility edge},\ }\href
  {https://doi.org/10.1103/PhysRevB.110.L041102} {\bibfield  {journal}
  {\bibinfo  {journal} {Phys. Rev. B}\ }\textbf {\bibinfo {volume} {110}},\
  \bibinfo {pages} {L041102} (\bibinfo {year} {2024})}\BibitemShut {NoStop}%
\bibitem [{\citenamefont {Wang}\ \emph
  {et~al.}(2024{\natexlab{b}})\citenamefont {Wang}, \citenamefont {Liu},
  \citenamefont {Wang},\ and\ \citenamefont {Chen}}]{flagellateprb}%
  \BibitemOpen
  \bibfield  {author} {\bibinfo {author} {\bibfnamefont {L.}~\bibnamefont
  {Wang}}, \bibinfo {author} {\bibfnamefont {J.}~\bibnamefont {Liu}}, \bibinfo
  {author} {\bibfnamefont {Z.}~\bibnamefont {Wang}},\ and\ \bibinfo {author}
  {\bibfnamefont {S.}~\bibnamefont {Chen}},\ }\bibfield  {title} {\bibinfo
  {title} {Exact complex mobility edges and flagellate-like spectra for
  non-hermitian quasicrystals with exponential hoppings},\ }\href
  {https://doi.org/10.1103/PhysRevB.110.144205} {\bibfield  {journal} {\bibinfo
   {journal} {Phys. Rev. B}\ }\textbf {\bibinfo {volume} {110}},\ \bibinfo
  {pages} {144205} (\bibinfo {year} {2024}{\natexlab{b}})}\BibitemShut
  {NoStop}%
\bibitem [{\citenamefont {Bordia}\ \emph {et~al.}(2016)\citenamefont {Bordia},
  \citenamefont {L\"uschen}, \citenamefont {Hodgman}, \citenamefont
  {Schreiber}, \citenamefont {Bloch},\ and\ \citenamefont
  {Schneider}}]{PhysRevLett.116.140401}%
  \BibitemOpen
  \bibfield  {author} {\bibinfo {author} {\bibfnamefont {P.}~\bibnamefont
  {Bordia}}, \bibinfo {author} {\bibfnamefont {H.~P.}\ \bibnamefont
  {L\"uschen}}, \bibinfo {author} {\bibfnamefont {S.~S.}\ \bibnamefont
  {Hodgman}}, \bibinfo {author} {\bibfnamefont {M.}~\bibnamefont {Schreiber}},
  \bibinfo {author} {\bibfnamefont {I.}~\bibnamefont {Bloch}},\ and\ \bibinfo
  {author} {\bibfnamefont {U.}~\bibnamefont {Schneider}},\ }\bibfield  {title}
  {\bibinfo {title} {Coupling identical one-dimensional many-body localized
  systems},\ }\href {https://doi.org/10.1103/PhysRevLett.116.140401} {\bibfield
   {journal} {\bibinfo  {journal} {Phys. Rev. Lett.}\ }\textbf {\bibinfo
  {volume} {116}},\ \bibinfo {pages} {140401} (\bibinfo {year}
  {2016})}\BibitemShut {NoStop}%
\bibitem [{\citenamefont {Bordia}\ \emph {et~al.}(2017)\citenamefont {Bordia},
  \citenamefont {L\"uschen}, \citenamefont {Scherg}, \citenamefont
  {Gopalakrishnan}, \citenamefont {Knap}, \citenamefont {Schneider},\ and\
  \citenamefont {Bloch}}]{PhysRevX.7.041047}%
  \BibitemOpen
  \bibfield  {author} {\bibinfo {author} {\bibfnamefont {P.}~\bibnamefont
  {Bordia}}, \bibinfo {author} {\bibfnamefont {H.}~\bibnamefont {L\"uschen}},
  \bibinfo {author} {\bibfnamefont {S.}~\bibnamefont {Scherg}}, \bibinfo
  {author} {\bibfnamefont {S.}~\bibnamefont {Gopalakrishnan}}, \bibinfo
  {author} {\bibfnamefont {M.}~\bibnamefont {Knap}}, \bibinfo {author}
  {\bibfnamefont {U.}~\bibnamefont {Schneider}},\ and\ \bibinfo {author}
  {\bibfnamefont {I.}~\bibnamefont {Bloch}},\ }\bibfield  {title} {\bibinfo
  {title} {Probing slow relaxation and many-body localization in
  two-dimensional quasiperiodic systems},\ }\href
  {https://doi.org/10.1103/PhysRevX.7.041047} {\bibfield  {journal} {\bibinfo
  {journal} {Phys. Rev. X}\ }\textbf {\bibinfo {volume} {7}},\ \bibinfo {pages}
  {041047} (\bibinfo {year} {2017})}\BibitemShut {NoStop}%
\bibitem [{\citenamefont {An}\ \emph {et~al.}(2021)\citenamefont {An},
  \citenamefont {Padavi\ifmmode~\acute{c}\else \'{c}\fi{}}, \citenamefont
  {Meier}, \citenamefont {Hegde}, \citenamefont {Ganeshan}, \citenamefont
  {Pixley}, \citenamefont {Vishveshwara},\ and\ \citenamefont
  {Gadway}}]{An21prl}%
  \BibitemOpen
  \bibfield  {author} {\bibinfo {author} {\bibfnamefont {F.~A.}\ \bibnamefont
  {An}}, \bibinfo {author} {\bibfnamefont {K.}~\bibnamefont
  {Padavi\ifmmode~\acute{c}\else \'{c}\fi{}}}, \bibinfo {author} {\bibfnamefont
  {E.~J.}\ \bibnamefont {Meier}}, \bibinfo {author} {\bibfnamefont
  {S.}~\bibnamefont {Hegde}}, \bibinfo {author} {\bibfnamefont
  {S.}~\bibnamefont {Ganeshan}}, \bibinfo {author} {\bibfnamefont {J.~H.}\
  \bibnamefont {Pixley}}, \bibinfo {author} {\bibfnamefont {S.}~\bibnamefont
  {Vishveshwara}},\ and\ \bibinfo {author} {\bibfnamefont {B.}~\bibnamefont
  {Gadway}},\ }\bibfield  {title} {\bibinfo {title} {Interactions and mobility
  edges: Observing the generalized aubry-andr\'e model},\ }\href
  {https://doi.org/10.1103/PhysRevLett.126.040603} {\bibfield  {journal}
  {\bibinfo  {journal} {Phys. Rev. Lett.}\ }\textbf {\bibinfo {volume} {126}},\
  \bibinfo {pages} {040603} (\bibinfo {year} {2021})}\BibitemShut {NoStop}%
\bibitem [{\citenamefont {Gao}\ \emph {et~al.}(2024)\citenamefont {Gao},
  \citenamefont {Khaymovich}, \citenamefont {Wang}, \citenamefont {Xu},
  \citenamefont {Iovan}, \citenamefont {Krishna}, \citenamefont {Jieensi},
  \citenamefont {Cataldo}, \citenamefont {Balatsky}, \citenamefont {Zwiller},\
  and\ \citenamefont {Elshaari}}]{GAO2024}%
  \BibitemOpen
  \bibfield  {author} {\bibinfo {author} {\bibfnamefont {J.}~\bibnamefont
  {Gao}}, \bibinfo {author} {\bibfnamefont {I.~M.}\ \bibnamefont {Khaymovich}},
  \bibinfo {author} {\bibfnamefont {X.-W.}\ \bibnamefont {Wang}}, \bibinfo
  {author} {\bibfnamefont {Z.-S.}\ \bibnamefont {Xu}}, \bibinfo {author}
  {\bibfnamefont {A.}~\bibnamefont {Iovan}}, \bibinfo {author} {\bibfnamefont
  {G.}~\bibnamefont {Krishna}}, \bibinfo {author} {\bibfnamefont
  {J.}~\bibnamefont {Jieensi}}, \bibinfo {author} {\bibfnamefont
  {A.}~\bibnamefont {Cataldo}}, \bibinfo {author} {\bibfnamefont {A.~V.}\
  \bibnamefont {Balatsky}}, \bibinfo {author} {\bibfnamefont {V.}~\bibnamefont
  {Zwiller}},\ and\ \bibinfo {author} {\bibfnamefont {A.~W.}\ \bibnamefont
  {Elshaari}},\ }\bibfield  {title} {\bibinfo {title} {Probing multi-mobility
  edges in quasiperiodic mosaic lattices},\ }\bibfield  {journal} {\bibinfo
  {journal} {Science Bulletin}\ }\href
  {https://doi.org/https://doi.org/10.1016/j.scib.2024.09.030}
  {https://doi.org/10.1016/j.scib.2024.09.030} (\bibinfo {year}
  {2024})\BibitemShut {NoStop}%
\bibitem [{\citenamefont {Roy}\ \emph {et~al.}(2021)\citenamefont {Roy},
  \citenamefont {Mishra}, \citenamefont {Tanatar},\ and\ \citenamefont
  {Basu}}]{Roy21prl}%
  \BibitemOpen
  \bibfield  {author} {\bibinfo {author} {\bibfnamefont {S.}~\bibnamefont
  {Roy}}, \bibinfo {author} {\bibfnamefont {T.}~\bibnamefont {Mishra}},
  \bibinfo {author} {\bibfnamefont {B.}~\bibnamefont {Tanatar}},\ and\ \bibinfo
  {author} {\bibfnamefont {S.}~\bibnamefont {Basu}},\ }\bibfield  {title}
  {\bibinfo {title} {Reentrant localization transition in a quasiperiodic
  chain},\ }\href {https://doi.org/10.1103/PhysRevLett.126.106803} {\bibfield
  {journal} {\bibinfo  {journal} {Phys. Rev. Lett.}\ }\textbf {\bibinfo
  {volume} {126}},\ \bibinfo {pages} {106803} (\bibinfo {year}
  {2021})}\BibitemShut {NoStop}%
\bibitem [{\citenamefont {Biddle}\ and\ \citenamefont
  {Das~Sarma}(2010)}]{Biddle10prl}%
  \BibitemOpen
  \bibfield  {author} {\bibinfo {author} {\bibfnamefont {J.}~\bibnamefont
  {Biddle}}\ and\ \bibinfo {author} {\bibfnamefont {S.}~\bibnamefont
  {Das~Sarma}},\ }\bibfield  {title} {\bibinfo {title} {Predicted mobility
  edges in one-dimensional incommensurate optical lattices: An exactly solvable
  model of anderson localization},\ }\href
  {https://doi.org/10.1103/PhysRevLett.104.070601} {\bibfield  {journal}
  {\bibinfo  {journal} {Phys. Rev. Lett.}\ }\textbf {\bibinfo {volume} {104}},\
  \bibinfo {pages} {070601} (\bibinfo {year} {2010})}\BibitemShut {NoStop}%
\bibitem [{\citenamefont {Biddle}\ \emph {et~al.}(2011)\citenamefont {Biddle},
  \citenamefont {Priour}, \citenamefont {Wang},\ and\ \citenamefont
  {Das~Sarma}}]{Biddle11prb}%
  \BibitemOpen
  \bibfield  {author} {\bibinfo {author} {\bibfnamefont {J.}~\bibnamefont
  {Biddle}}, \bibinfo {author} {\bibfnamefont {D.~J.}\ \bibnamefont {Priour}},
  \bibinfo {author} {\bibfnamefont {B.}~\bibnamefont {Wang}},\ and\ \bibinfo
  {author} {\bibfnamefont {S.}~\bibnamefont {Das~Sarma}},\ }\bibfield  {title}
  {\bibinfo {title} {Localization in one-dimensional lattices with
  non-nearest-neighbor hopping: Generalized anderson and aubry-andr\'e
  models},\ }\href {https://doi.org/10.1103/PhysRevB.83.075105} {\bibfield
  {journal} {\bibinfo  {journal} {Phys. Rev. B}\ }\textbf {\bibinfo {volume}
  {83}},\ \bibinfo {pages} {075105} (\bibinfo {year} {2011})}\BibitemShut
  {NoStop}%
\bibitem [{\citenamefont {Deng}\ \emph {et~al.}(2019)\citenamefont {Deng},
  \citenamefont {Ray}, \citenamefont {Sinha}, \citenamefont {Shlyapnikov},\
  and\ \citenamefont {Santos}}]{Santos19prl}%
  \BibitemOpen
  \bibfield  {author} {\bibinfo {author} {\bibfnamefont {X.}~\bibnamefont
  {Deng}}, \bibinfo {author} {\bibfnamefont {S.}~\bibnamefont {Ray}}, \bibinfo
  {author} {\bibfnamefont {S.}~\bibnamefont {Sinha}}, \bibinfo {author}
  {\bibfnamefont {G.~V.}\ \bibnamefont {Shlyapnikov}},\ and\ \bibinfo {author}
  {\bibfnamefont {L.}~\bibnamefont {Santos}},\ }\bibfield  {title} {\bibinfo
  {title} {One-dimensional quasicrystals with power-law hopping},\ }\href
  {https://doi.org/10.1103/PhysRevLett.123.025301} {\bibfield  {journal}
  {\bibinfo  {journal} {Phys. Rev. Lett.}\ }\textbf {\bibinfo {volume} {123}},\
  \bibinfo {pages} {025301} (\bibinfo {year} {2019})}\BibitemShut {NoStop}%
\bibitem [{\citenamefont {Roy}\ and\ \citenamefont
  {Sharma}(2021)}]{prb103075124}%
  \BibitemOpen
  \bibfield  {author} {\bibinfo {author} {\bibfnamefont {N.}~\bibnamefont
  {Roy}}\ and\ \bibinfo {author} {\bibfnamefont {A.}~\bibnamefont {Sharma}},\
  }\bibfield  {title} {\bibinfo {title} {Fraction of delocalized eigenstates in
  the long-range aubry-andr\'e-harper model},\ }\href
  {https://doi.org/10.1103/PhysRevB.103.075124} {\bibfield  {journal} {\bibinfo
   {journal} {Phys. Rev. B}\ }\textbf {\bibinfo {volume} {103}},\ \bibinfo
  {pages} {075124} (\bibinfo {year} {2021})}\BibitemShut {NoStop}%
\bibitem [{\citenamefont {Das~Sarma}\ \emph {et~al.}(1988)\citenamefont
  {Das~Sarma}, \citenamefont {He},\ and\ \citenamefont {Xie}}]{xiexc1988prl}%
  \BibitemOpen
  \bibfield  {author} {\bibinfo {author} {\bibfnamefont {S.}~\bibnamefont
  {Das~Sarma}}, \bibinfo {author} {\bibfnamefont {S.}~\bibnamefont {He}},\ and\
  \bibinfo {author} {\bibfnamefont {X.~C.}\ \bibnamefont {Xie}},\ }\bibfield
  {title} {\bibinfo {title} {Mobility edge in a model one-dimensional
  potential},\ }\href {https://doi.org/10.1103/PhysRevLett.61.2144} {\bibfield
  {journal} {\bibinfo  {journal} {Phys. Rev. Lett.}\ }\textbf {\bibinfo
  {volume} {61}},\ \bibinfo {pages} {2144} (\bibinfo {year}
  {1988})}\BibitemShut {NoStop}%
\bibitem [{\citenamefont {Das~Sarma}\ \emph {et~al.}(1990)\citenamefont
  {Das~Sarma}, \citenamefont {He},\ and\ \citenamefont {Xie}}]{xieprb415544}%
  \BibitemOpen
  \bibfield  {author} {\bibinfo {author} {\bibfnamefont {S.}~\bibnamefont
  {Das~Sarma}}, \bibinfo {author} {\bibfnamefont {S.}~\bibnamefont {He}},\ and\
  \bibinfo {author} {\bibfnamefont {X.~C.}\ \bibnamefont {Xie}},\ }\bibfield
  {title} {\bibinfo {title} {Localization, mobility edges, and metal-insulator
  transition in a class of one-dimensional slowly varying deterministic
  potentials},\ }\href {https://doi.org/10.1103/PhysRevB.41.5544} {\bibfield
  {journal} {\bibinfo  {journal} {Phys. Rev. B}\ }\textbf {\bibinfo {volume}
  {41}},\ \bibinfo {pages} {5544} (\bibinfo {year} {1990})}\BibitemShut
  {NoStop}%
\bibitem [{\citenamefont {Ganeshan}\ \emph {et~al.}(2015)\citenamefont
  {Ganeshan}, \citenamefont {Pixley},\ and\ \citenamefont
  {Das~Sarma}}]{Ganeshan2015prl}%
  \BibitemOpen
  \bibfield  {author} {\bibinfo {author} {\bibfnamefont {S.}~\bibnamefont
  {Ganeshan}}, \bibinfo {author} {\bibfnamefont {J.~H.}\ \bibnamefont
  {Pixley}},\ and\ \bibinfo {author} {\bibfnamefont {S.}~\bibnamefont
  {Das~Sarma}},\ }\bibfield  {title} {\bibinfo {title} {Nearest neighbor tight
  binding models with an exact mobility edge in one dimension},\ }\href
  {https://doi.org/10.1103/PhysRevLett.114.146601} {\bibfield  {journal}
  {\bibinfo  {journal} {Phys. Rev. Lett.}\ }\textbf {\bibinfo {volume} {114}},\
  \bibinfo {pages} {146601} (\bibinfo {year} {2015})}\BibitemShut {NoStop}%
\bibitem [{\citenamefont {Li}\ \emph {et~al.}(2016)\citenamefont {Li},
  \citenamefont {Pixley}, \citenamefont {Deng}, \citenamefont {Ganeshan},\ and\
  \citenamefont {Das~Sarma}}]{Lixp16prb}%
  \BibitemOpen
  \bibfield  {author} {\bibinfo {author} {\bibfnamefont {X.}~\bibnamefont
  {Li}}, \bibinfo {author} {\bibfnamefont {J.~H.}\ \bibnamefont {Pixley}},
  \bibinfo {author} {\bibfnamefont {D.-L.}\ \bibnamefont {Deng}}, \bibinfo
  {author} {\bibfnamefont {S.}~\bibnamefont {Ganeshan}},\ and\ \bibinfo
  {author} {\bibfnamefont {S.}~\bibnamefont {Das~Sarma}},\ }\bibfield  {title}
  {\bibinfo {title} {Quantum nonergodicity and fermion localization in a system
  with a single-particle mobility edge},\ }\href
  {https://doi.org/10.1103/PhysRevB.93.184204} {\bibfield  {journal} {\bibinfo
  {journal} {Phys. Rev. B}\ }\textbf {\bibinfo {volume} {93}},\ \bibinfo
  {pages} {184204} (\bibinfo {year} {2016})}\BibitemShut {NoStop}%
\bibitem [{\citenamefont {Li}\ \emph {et~al.}(2017)\citenamefont {Li},
  \citenamefont {Li},\ and\ \citenamefont {Das~Sarma}}]{Sarma17prb}%
  \BibitemOpen
  \bibfield  {author} {\bibinfo {author} {\bibfnamefont {X.}~\bibnamefont
  {Li}}, \bibinfo {author} {\bibfnamefont {X.}~\bibnamefont {Li}},\ and\
  \bibinfo {author} {\bibfnamefont {S.}~\bibnamefont {Das~Sarma}},\ }\bibfield
  {title} {\bibinfo {title} {Mobility edges in one-dimensional bichromatic
  incommensurate potentials},\ }\href
  {https://doi.org/10.1103/PhysRevB.96.085119} {\bibfield  {journal} {\bibinfo
  {journal} {Phys. Rev. B}\ }\textbf {\bibinfo {volume} {96}},\ \bibinfo
  {pages} {085119} (\bibinfo {year} {2017})}\BibitemShut {NoStop}%
\bibitem [{\citenamefont {Li}\ and\ \citenamefont
  {Das~Sarma}(2020)}]{Lix20prb}%
  \BibitemOpen
  \bibfield  {author} {\bibinfo {author} {\bibfnamefont {X.}~\bibnamefont
  {Li}}\ and\ \bibinfo {author} {\bibfnamefont {S.}~\bibnamefont {Das~Sarma}},\
  }\bibfield  {title} {\bibinfo {title} {Mobility edge and intermediate phase
  in one-dimensional incommensurate lattice potentials},\ }\href
  {https://doi.org/10.1103/PhysRevB.101.064203} {\bibfield  {journal} {\bibinfo
   {journal} {Phys. Rev. B}\ }\textbf {\bibinfo {volume} {101}},\ \bibinfo
  {pages} {064203} (\bibinfo {year} {2020})}\BibitemShut {NoStop}%
\bibitem [{\citenamefont {Padhan}\ \emph {et~al.}(2022)\citenamefont {Padhan},
  \citenamefont {Giri}, \citenamefont {Mondal},\ and\ \citenamefont
  {Mishra}}]{prb105L220201}%
  \BibitemOpen
  \bibfield  {author} {\bibinfo {author} {\bibfnamefont {A.}~\bibnamefont
  {Padhan}}, \bibinfo {author} {\bibfnamefont {M.~K.}\ \bibnamefont {Giri}},
  \bibinfo {author} {\bibfnamefont {S.}~\bibnamefont {Mondal}},\ and\ \bibinfo
  {author} {\bibfnamefont {T.}~\bibnamefont {Mishra}},\ }\bibfield  {title}
  {\bibinfo {title} {Emergence of multiple localization transitions in a
  one-dimensional quasiperiodic lattice},\ }\href
  {https://doi.org/10.1103/PhysRevB.105.L220201} {\bibfield  {journal}
  {\bibinfo  {journal} {Phys. Rev. B}\ }\textbf {\bibinfo {volume} {105}},\
  \bibinfo {pages} {L220201} (\bibinfo {year} {2022})}\BibitemShut {NoStop}%
\bibitem [{\citenamefont {Liu}\ \emph {et~al.}(2017)\citenamefont {Liu},
  \citenamefont {Yan},\ and\ \citenamefont {Guo}}]{prb96174207}%
  \BibitemOpen
  \bibfield  {author} {\bibinfo {author} {\bibfnamefont {T.}~\bibnamefont
  {Liu}}, \bibinfo {author} {\bibfnamefont {H.-Y.}\ \bibnamefont {Yan}},\ and\
  \bibinfo {author} {\bibfnamefont {H.}~\bibnamefont {Guo}},\ }\bibfield
  {title} {\bibinfo {title} {Fate of topological states and mobility edges in
  one-dimensional slowly varying incommensurate potentials},\ }\href
  {https://doi.org/10.1103/PhysRevB.96.174207} {\bibfield  {journal} {\bibinfo
  {journal} {Phys. Rev. B}\ }\textbf {\bibinfo {volume} {96}},\ \bibinfo
  {pages} {174207} (\bibinfo {year} {2017})}\BibitemShut {NoStop}%
\bibitem [{\citenamefont {Bender}\ and\ \citenamefont
  {Boettcher}(1998)}]{bender98}%
  \BibitemOpen
  \bibfield  {author} {\bibinfo {author} {\bibfnamefont {C.~M.}\ \bibnamefont
  {Bender}}\ and\ \bibinfo {author} {\bibfnamefont {S.}~\bibnamefont
  {Boettcher}},\ }\bibfield  {title} {\bibinfo {title} {Real spectra in
  non-hermitian hamiltonians having $\mathcal{PT}$ symmetry},\ }\href
  {https://doi.org/10.1103/PhysRevLett.80.5243} {\bibfield  {journal} {\bibinfo
   {journal} {Phys. Rev. Lett.}\ }\textbf {\bibinfo {volume} {80}},\ \bibinfo
  {pages} {5243} (\bibinfo {year} {1998})}\BibitemShut {NoStop}%
\bibitem [{\citenamefont {Bender}(2007)}]{RPP70947}%
  \BibitemOpen
  \bibfield  {author} {\bibinfo {author} {\bibfnamefont {C.~M.}\ \bibnamefont
  {Bender}},\ }\bibfield  {title} {\bibinfo {title} {Making sense of
  non-hermitian hamiltonians},\ }\href
  {https://doi.org/10.1088/0034-4885/70/6/R03} {\bibfield  {journal} {\bibinfo
  {journal} {Reports on Progress in Physics}\ }\textbf {\bibinfo {volume}
  {70}},\ \bibinfo {pages} {947} (\bibinfo {year} {2007})}\BibitemShut
  {NoStop}%
\bibitem [{\citenamefont {El-Ganainy}\ \emph {et~al.}(2018)\citenamefont
  {El-Ganainy}, \citenamefont {Makris}, \citenamefont {Khajavikhan},
  \citenamefont {Musslimani}, \citenamefont {Rotter},\ and\ \citenamefont
  {Christodoulides}}]{NP1411}%
  \BibitemOpen
  \bibfield  {author} {\bibinfo {author} {\bibfnamefont {R.}~\bibnamefont
  {El-Ganainy}}, \bibinfo {author} {\bibfnamefont {K.~G.}\ \bibnamefont
  {Makris}}, \bibinfo {author} {\bibfnamefont {M.}~\bibnamefont {Khajavikhan}},
  \bibinfo {author} {\bibfnamefont {Z.~H.}\ \bibnamefont {Musslimani}},
  \bibinfo {author} {\bibfnamefont {S.}~\bibnamefont {Rotter}},\ and\ \bibinfo
  {author} {\bibfnamefont {D.~N.}\ \bibnamefont {Christodoulides}},\ }\bibfield
   {title} {\bibinfo {title} {Non-hermitian physics and
  $\mathcal{P}\mathcal{T}$ symmetry},\ }\href
  {https://doi.org/10.1038/nphys4323} {\bibfield  {journal} {\bibinfo
  {journal} {Nature Physics}\ }\textbf {\bibinfo {volume} {14}},\ \bibinfo
  {pages} {11} (\bibinfo {year} {2018})}\BibitemShut {NoStop}%
\bibitem [{\citenamefont {Yuto~Ashida}\ and\ \citenamefont
  {Ueda}(2020)}]{AP69249}%
  \BibitemOpen
  \bibfield  {author} {\bibinfo {author} {\bibfnamefont {Z.~G.}\ \bibnamefont
  {Yuto~Ashida}}\ and\ \bibinfo {author} {\bibfnamefont {M.}~\bibnamefont
  {Ueda}},\ }\bibfield  {title} {\bibinfo {title} {Non-hermitian physics},\
  }\href {https://doi.org/10.1080/00018732.2021.1876991} {\bibfield  {journal}
  {\bibinfo  {journal} {Advances in Physics}\ }\textbf {\bibinfo {volume}
  {69}},\ \bibinfo {pages} {249} (\bibinfo {year} {2020})},\ \Eprint
  {https://arxiv.org/abs/https://doi.org/10.1080/00018732.2021.1876991}
  {https://doi.org/10.1080/00018732.2021.1876991} \BibitemShut {NoStop}%
\bibitem [{\citenamefont {Xiao}\ \emph {et~al.}(2017)\citenamefont {Xiao},
  \citenamefont {Zhan}, \citenamefont {Bian}, \citenamefont {Wang},
  \citenamefont {Zhang}, \citenamefont {Wang}, \citenamefont {Li},
  \citenamefont {Mochizuki}, \citenamefont {Kim}, \citenamefont {Kawakami},
  \citenamefont {Yi}, \citenamefont {Obuse}, \citenamefont {Sanders},\ and\
  \citenamefont {Xue}}]{NP131117}%
  \BibitemOpen
  \bibfield  {author} {\bibinfo {author} {\bibfnamefont {L.}~\bibnamefont
  {Xiao}}, \bibinfo {author} {\bibfnamefont {X.}~\bibnamefont {Zhan}}, \bibinfo
  {author} {\bibfnamefont {Z.~H.}\ \bibnamefont {Bian}}, \bibinfo {author}
  {\bibfnamefont {K.~K.}\ \bibnamefont {Wang}}, \bibinfo {author}
  {\bibfnamefont {X.}~\bibnamefont {Zhang}}, \bibinfo {author} {\bibfnamefont
  {X.~P.}\ \bibnamefont {Wang}}, \bibinfo {author} {\bibfnamefont
  {J.}~\bibnamefont {Li}}, \bibinfo {author} {\bibfnamefont {K.}~\bibnamefont
  {Mochizuki}}, \bibinfo {author} {\bibfnamefont {D.}~\bibnamefont {Kim}},
  \bibinfo {author} {\bibfnamefont {N.}~\bibnamefont {Kawakami}}, \bibinfo
  {author} {\bibfnamefont {W.}~\bibnamefont {Yi}}, \bibinfo {author}
  {\bibfnamefont {H.}~\bibnamefont {Obuse}}, \bibinfo {author} {\bibfnamefont
  {B.~C.}\ \bibnamefont {Sanders}},\ and\ \bibinfo {author} {\bibfnamefont
  {P.}~\bibnamefont {Xue}},\ }\bibfield  {title} {\bibinfo {title} {Observation
  of topological edge states in parity-time-symmetric quantum walks},\ }\href
  {https://doi.org/10.1038/nphys4204} {\bibfield  {journal} {\bibinfo
  {journal} {Nature Physics}\ }\textbf {\bibinfo {volume} {13}},\ \bibinfo
  {pages} {1117} (\bibinfo {year} {2017})}\BibitemShut {NoStop}%
\bibitem [{\citenamefont {Yao}\ and\ \citenamefont
  {Wang}(2018)}]{prl121086803}%
  \BibitemOpen
  \bibfield  {author} {\bibinfo {author} {\bibfnamefont {S.}~\bibnamefont
  {Yao}}\ and\ \bibinfo {author} {\bibfnamefont {Z.}~\bibnamefont {Wang}},\
  }\bibfield  {title} {\bibinfo {title} {Edge states and topological invariants
  of non-hermitian systems},\ }\href
  {https://doi.org/10.1103/PhysRevLett.121.086803} {\bibfield  {journal}
  {\bibinfo  {journal} {Phys. Rev. Lett.}\ }\textbf {\bibinfo {volume} {121}},\
  \bibinfo {pages} {086803} (\bibinfo {year} {2018})}\BibitemShut {NoStop}%
\bibitem [{\citenamefont {Kunst}\ \emph {et~al.}(2018)\citenamefont {Kunst},
  \citenamefont {Edvardsson}, \citenamefont {Budich},\ and\ \citenamefont
  {Bergholtz}}]{prl121.026808}%
  \BibitemOpen
  \bibfield  {author} {\bibinfo {author} {\bibfnamefont {F.~K.}\ \bibnamefont
  {Kunst}}, \bibinfo {author} {\bibfnamefont {E.}~\bibnamefont {Edvardsson}},
  \bibinfo {author} {\bibfnamefont {J.~C.}\ \bibnamefont {Budich}},\ and\
  \bibinfo {author} {\bibfnamefont {E.~J.}\ \bibnamefont {Bergholtz}},\
  }\bibfield  {title} {\bibinfo {title} {Biorthogonal bulk-boundary
  correspondence in non-hermitian systems},\ }\href
  {https://doi.org/10.1103/PhysRevLett.121.026808} {\bibfield  {journal}
  {\bibinfo  {journal} {Phys. Rev. Lett.}\ }\textbf {\bibinfo {volume} {121}},\
  \bibinfo {pages} {026808} (\bibinfo {year} {2018})}\BibitemShut {NoStop}%
\bibitem [{\citenamefont {Xiong}(2018)}]{JPC2035043}%
  \BibitemOpen
  \bibfield  {author} {\bibinfo {author} {\bibfnamefont {Y.}~\bibnamefont
  {Xiong}},\ }\bibfield  {title} {\bibinfo {title} {Why does bulk boundary
  correspondence fail in some non-hermitian topological models},\ }\href
  {https://doi.org/10.1088/2399-6528/aab64a} {\bibfield  {journal} {\bibinfo
  {journal} {Journal of Physics Communications}\ }\textbf {\bibinfo {volume}
  {2}},\ \bibinfo {pages} {035043} (\bibinfo {year} {2018})}\BibitemShut
  {NoStop}%
\bibitem [{\citenamefont {Gong}\ \emph {et~al.}(2018)\citenamefont {Gong},
  \citenamefont {Ashida}, \citenamefont {Kawabata}, \citenamefont {Takasan},
  \citenamefont {Higashikawa},\ and\ \citenamefont {Ueda}}]{PRX8031079}%
  \BibitemOpen
  \bibfield  {author} {\bibinfo {author} {\bibfnamefont {Z.}~\bibnamefont
  {Gong}}, \bibinfo {author} {\bibfnamefont {Y.}~\bibnamefont {Ashida}},
  \bibinfo {author} {\bibfnamefont {K.}~\bibnamefont {Kawabata}}, \bibinfo
  {author} {\bibfnamefont {K.}~\bibnamefont {Takasan}}, \bibinfo {author}
  {\bibfnamefont {S.}~\bibnamefont {Higashikawa}},\ and\ \bibinfo {author}
  {\bibfnamefont {M.}~\bibnamefont {Ueda}},\ }\bibfield  {title} {\bibinfo
  {title} {Topological phases of non-hermitian systems},\ }\href
  {https://doi.org/10.1103/PhysRevX.8.031079} {\bibfield  {journal} {\bibinfo
  {journal} {Phys. Rev. X}\ }\textbf {\bibinfo {volume} {8}},\ \bibinfo {pages}
  {031079} (\bibinfo {year} {2018})}\BibitemShut {NoStop}%
\bibitem [{\citenamefont {Kawabata}\ \emph {et~al.}(2019)\citenamefont
  {Kawabata}, \citenamefont {Shiozaki}, \citenamefont {Ueda},\ and\
  \citenamefont {Sato}}]{PRX9041015}%
  \BibitemOpen
  \bibfield  {author} {\bibinfo {author} {\bibfnamefont {K.}~\bibnamefont
  {Kawabata}}, \bibinfo {author} {\bibfnamefont {K.}~\bibnamefont {Shiozaki}},
  \bibinfo {author} {\bibfnamefont {M.}~\bibnamefont {Ueda}},\ and\ \bibinfo
  {author} {\bibfnamefont {M.}~\bibnamefont {Sato}},\ }\bibfield  {title}
  {\bibinfo {title} {Symmetry and topology in non-hermitian physics},\ }\href
  {https://doi.org/10.1103/PhysRevX.9.041015} {\bibfield  {journal} {\bibinfo
  {journal} {Phys. Rev. X}\ }\textbf {\bibinfo {volume} {9}},\ \bibinfo {pages}
  {041015} (\bibinfo {year} {2019})}\BibitemShut {NoStop}%
\bibitem [{\citenamefont {Lee}\ and\ \citenamefont
  {Thomale}(2019)}]{PRB99201103}%
  \BibitemOpen
  \bibfield  {author} {\bibinfo {author} {\bibfnamefont {C.~H.}\ \bibnamefont
  {Lee}}\ and\ \bibinfo {author} {\bibfnamefont {R.}~\bibnamefont {Thomale}},\
  }\bibfield  {title} {\bibinfo {title} {Anatomy of skin modes and topology in
  non-hermitian systems},\ }\href {https://doi.org/10.1103/PhysRevB.99.201103}
  {\bibfield  {journal} {\bibinfo  {journal} {Phys. Rev. B}\ }\textbf {\bibinfo
  {volume} {99}},\ \bibinfo {pages} {201103} (\bibinfo {year}
  {2019})}\BibitemShut {NoStop}%
\bibitem [{\citenamefont {Yokomizo}\ and\ \citenamefont
  {Murakami}(2019)}]{PRL123066404}%
  \BibitemOpen
  \bibfield  {author} {\bibinfo {author} {\bibfnamefont {K.}~\bibnamefont
  {Yokomizo}}\ and\ \bibinfo {author} {\bibfnamefont {S.}~\bibnamefont
  {Murakami}},\ }\bibfield  {title} {\bibinfo {title} {Non-bloch band theory of
  non-hermitian systems},\ }\href
  {https://doi.org/10.1103/PhysRevLett.123.066404} {\bibfield  {journal}
  {\bibinfo  {journal} {Phys. Rev. Lett.}\ }\textbf {\bibinfo {volume} {123}},\
  \bibinfo {pages} {066404} (\bibinfo {year} {2019})}\BibitemShut {NoStop}%
\bibitem [{\citenamefont {Xiao}\ \emph {et~al.}(2020)\citenamefont {Xiao},
  \citenamefont {Deng}, \citenamefont {Wang}, \citenamefont {Zhu},
  \citenamefont {Wang}, \citenamefont {Yi},\ and\ \citenamefont
  {Xue}}]{NP16761}%
  \BibitemOpen
  \bibfield  {author} {\bibinfo {author} {\bibfnamefont {L.}~\bibnamefont
  {Xiao}}, \bibinfo {author} {\bibfnamefont {T.}~\bibnamefont {Deng}}, \bibinfo
  {author} {\bibfnamefont {K.}~\bibnamefont {Wang}}, \bibinfo {author}
  {\bibfnamefont {G.}~\bibnamefont {Zhu}}, \bibinfo {author} {\bibfnamefont
  {Z.}~\bibnamefont {Wang}}, \bibinfo {author} {\bibfnamefont {W.}~\bibnamefont
  {Yi}},\ and\ \bibinfo {author} {\bibfnamefont {P.}~\bibnamefont {Xue}},\
  }\bibfield  {title} {\bibinfo {title} {Non-hermitian bulk-boundary
  correspondence in quantum dynamics},\ }\href
  {https://doi.org/10.1038/s41567-020-0836-6} {\bibfield  {journal} {\bibinfo
  {journal} {Nature Physics}\ }\textbf {\bibinfo {volume} {16}},\ \bibinfo
  {pages} {761} (\bibinfo {year} {2020})}\BibitemShut {NoStop}%
\bibitem [{\citenamefont {Zhang}\ \emph {et~al.}(2020)\citenamefont {Zhang},
  \citenamefont {Yang},\ and\ \citenamefont {Fang}}]{PRL125126402}%
  \BibitemOpen
  \bibfield  {author} {\bibinfo {author} {\bibfnamefont {K.}~\bibnamefont
  {Zhang}}, \bibinfo {author} {\bibfnamefont {Z.}~\bibnamefont {Yang}},\ and\
  \bibinfo {author} {\bibfnamefont {C.}~\bibnamefont {Fang}},\ }\bibfield
  {title} {\bibinfo {title} {Correspondence between winding numbers and skin
  modes in non-hermitian systems},\ }\href
  {https://doi.org/10.1103/PhysRevLett.125.126402} {\bibfield  {journal}
  {\bibinfo  {journal} {Phys. Rev. Lett.}\ }\textbf {\bibinfo {volume} {125}},\
  \bibinfo {pages} {126402} (\bibinfo {year} {2020})}\BibitemShut {NoStop}%
\bibitem [{\citenamefont {Okuma}\ \emph {et~al.}(2020)\citenamefont {Okuma},
  \citenamefont {Kawabata}, \citenamefont {Shiozaki},\ and\ \citenamefont
  {Sato}}]{PRL124086801}%
  \BibitemOpen
  \bibfield  {author} {\bibinfo {author} {\bibfnamefont {N.}~\bibnamefont
  {Okuma}}, \bibinfo {author} {\bibfnamefont {K.}~\bibnamefont {Kawabata}},
  \bibinfo {author} {\bibfnamefont {K.}~\bibnamefont {Shiozaki}},\ and\
  \bibinfo {author} {\bibfnamefont {M.}~\bibnamefont {Sato}},\ }\bibfield
  {title} {\bibinfo {title} {Topological origin of non-hermitian skin
  effects},\ }\href {https://doi.org/10.1103/PhysRevLett.124.086801} {\bibfield
   {journal} {\bibinfo  {journal} {Phys. Rev. Lett.}\ }\textbf {\bibinfo
  {volume} {124}},\ \bibinfo {pages} {086801} (\bibinfo {year}
  {2020})}\BibitemShut {NoStop}%
\bibitem [{\citenamefont {Wang}\ \emph {et~al.}(2021)\citenamefont {Wang},
  \citenamefont {Liu},\ and\ \citenamefont {Zhang}}]{CPB30020506}%
  \BibitemOpen
  \bibfield  {author} {\bibinfo {author} {\bibfnamefont {L.}~\bibnamefont
  {Wang}}, \bibinfo {author} {\bibfnamefont {Q.}~\bibnamefont {Liu}},\ and\
  \bibinfo {author} {\bibfnamefont {Y.}~\bibnamefont {Zhang}},\ }\bibfield
  {title} {\bibinfo {title} {Quantum dynamics on a lossy non-hermitian
  lattice},\ }\href {https://doi.org/10.1088/1674-1056/abd765} {\bibfield
  {journal} {\bibinfo  {journal} {Chinese Physics B}\ }\textbf {\bibinfo
  {volume} {30}},\ \bibinfo {pages} {020506} (\bibinfo {year}
  {2021})}\BibitemShut {NoStop}%
\bibitem [{\citenamefont {Xue}\ \emph {et~al.}(2022)\citenamefont {Xue},
  \citenamefont {Hu}, \citenamefont {Song},\ and\ \citenamefont
  {Wang}}]{prl128120401}%
  \BibitemOpen
  \bibfield  {author} {\bibinfo {author} {\bibfnamefont {W.-T.}\ \bibnamefont
  {Xue}}, \bibinfo {author} {\bibfnamefont {Y.-M.}\ \bibnamefont {Hu}},
  \bibinfo {author} {\bibfnamefont {F.}~\bibnamefont {Song}},\ and\ \bibinfo
  {author} {\bibfnamefont {Z.}~\bibnamefont {Wang}},\ }\bibfield  {title}
  {\bibinfo {title} {Non-hermitian edge burst},\ }\href
  {https://doi.org/10.1103/PhysRevLett.128.120401} {\bibfield  {journal}
  {\bibinfo  {journal} {Phys. Rev. Lett.}\ }\textbf {\bibinfo {volume} {128}},\
  \bibinfo {pages} {120401} (\bibinfo {year} {2022})}\BibitemShut {NoStop}%
\bibitem [{\citenamefont {Borgnia}\ \emph {et~al.}(2020)\citenamefont
  {Borgnia}, \citenamefont {Kruchkov},\ and\ \citenamefont
  {Slager}}]{PRL124056802}%
  \BibitemOpen
  \bibfield  {author} {\bibinfo {author} {\bibfnamefont {D.~S.}\ \bibnamefont
  {Borgnia}}, \bibinfo {author} {\bibfnamefont {A.~J.}\ \bibnamefont
  {Kruchkov}},\ and\ \bibinfo {author} {\bibfnamefont {R.-J.}\ \bibnamefont
  {Slager}},\ }\bibfield  {title} {\bibinfo {title} {Non-hermitian boundary
  modes and topology},\ }\href {https://doi.org/10.1103/PhysRevLett.124.056802}
  {\bibfield  {journal} {\bibinfo  {journal} {Phys. Rev. Lett.}\ }\textbf
  {\bibinfo {volume} {124}},\ \bibinfo {pages} {056802} (\bibinfo {year}
  {2020})}\BibitemShut {NoStop}%
\bibitem [{\citenamefont {Yang}\ \emph {et~al.}(2020)\citenamefont {Yang},
  \citenamefont {Zhang}, \citenamefont {Fang},\ and\ \citenamefont
  {Hu}}]{PRL125226402}%
  \BibitemOpen
  \bibfield  {author} {\bibinfo {author} {\bibfnamefont {Z.}~\bibnamefont
  {Yang}}, \bibinfo {author} {\bibfnamefont {K.}~\bibnamefont {Zhang}},
  \bibinfo {author} {\bibfnamefont {C.}~\bibnamefont {Fang}},\ and\ \bibinfo
  {author} {\bibfnamefont {J.}~\bibnamefont {Hu}},\ }\bibfield  {title}
  {\bibinfo {title} {Non-hermitian bulk-boundary correspondence and auxiliary
  generalized brillouin zone theory},\ }\href
  {https://doi.org/10.1103/PhysRevLett.125.226402} {\bibfield  {journal}
  {\bibinfo  {journal} {Phys. Rev. Lett.}\ }\textbf {\bibinfo {volume} {125}},\
  \bibinfo {pages} {226402} (\bibinfo {year} {2020})}\BibitemShut {NoStop}%
\bibitem [{\citenamefont {Guo}\ \emph {et~al.}(2021)\citenamefont {Guo},
  \citenamefont {Liu}, \citenamefont {Zhao}, \citenamefont {Liu},\ and\
  \citenamefont {Chen}}]{PRL127116801}%
  \BibitemOpen
  \bibfield  {author} {\bibinfo {author} {\bibfnamefont {C.-X.}\ \bibnamefont
  {Guo}}, \bibinfo {author} {\bibfnamefont {C.-H.}\ \bibnamefont {Liu}},
  \bibinfo {author} {\bibfnamefont {X.-M.}\ \bibnamefont {Zhao}}, \bibinfo
  {author} {\bibfnamefont {Y.}~\bibnamefont {Liu}},\ and\ \bibinfo {author}
  {\bibfnamefont {S.}~\bibnamefont {Chen}},\ }\bibfield  {title} {\bibinfo
  {title} {Exact solution of non-hermitian systems with generalized boundary
  conditions: Size-dependent boundary effect and fragility of the skin
  effect},\ }\href {https://doi.org/10.1103/PhysRevLett.127.116801} {\bibfield
  {journal} {\bibinfo  {journal} {Phys. Rev. Lett.}\ }\textbf {\bibinfo
  {volume} {127}},\ \bibinfo {pages} {116801} (\bibinfo {year}
  {2021})}\BibitemShut {NoStop}%
\bibitem [{\citenamefont {Li}\ \emph {et~al.}(2020)\citenamefont {Li},
  \citenamefont {Lee}, \citenamefont {Mu},\ and\ \citenamefont
  {Gong}}]{NC115491}%
  \BibitemOpen
  \bibfield  {author} {\bibinfo {author} {\bibfnamefont {L.}~\bibnamefont
  {Li}}, \bibinfo {author} {\bibfnamefont {C.~H.}\ \bibnamefont {Lee}},
  \bibinfo {author} {\bibfnamefont {S.}~\bibnamefont {Mu}},\ and\ \bibinfo
  {author} {\bibfnamefont {J.}~\bibnamefont {Gong}},\ }\bibfield  {title}
  {\bibinfo {title} {Critical non-hermitian skin effect},\ }\href
  {https://doi.org/10.1038/s41467-020-18917-4} {\bibfield  {journal} {\bibinfo
  {journal} {Nature Communications}\ }\textbf {\bibinfo {volume} {11}},\
  \bibinfo {pages} {5491} (\bibinfo {year} {2020})}\BibitemShut {NoStop}%
\bibitem [{\citenamefont {Bergholtz}\ \emph {et~al.}(2021)\citenamefont
  {Bergholtz}, \citenamefont {Budich},\ and\ \citenamefont
  {Kunst}}]{RMP93015005}%
  \BibitemOpen
  \bibfield  {author} {\bibinfo {author} {\bibfnamefont {E.~J.}\ \bibnamefont
  {Bergholtz}}, \bibinfo {author} {\bibfnamefont {J.~C.}\ \bibnamefont
  {Budich}},\ and\ \bibinfo {author} {\bibfnamefont {F.~K.}\ \bibnamefont
  {Kunst}},\ }\bibfield  {title} {\bibinfo {title} {Exceptional topology of
  non-hermitian systems},\ }\href
  {https://doi.org/10.1103/RevModPhys.93.015005} {\bibfield  {journal}
  {\bibinfo  {journal} {Rev. Mod. Phys.}\ }\textbf {\bibinfo {volume} {93}},\
  \bibinfo {pages} {015005} (\bibinfo {year} {2021})}\BibitemShut {NoStop}%
\bibitem [{\citenamefont {Kawabata}\ \emph {et~al.}(2023)\citenamefont
  {Kawabata}, \citenamefont {Numasawa},\ and\ \citenamefont
  {Ryu}}]{PRX13021007}%
  \BibitemOpen
  \bibfield  {author} {\bibinfo {author} {\bibfnamefont {K.}~\bibnamefont
  {Kawabata}}, \bibinfo {author} {\bibfnamefont {T.}~\bibnamefont {Numasawa}},\
  and\ \bibinfo {author} {\bibfnamefont {S.}~\bibnamefont {Ryu}},\ }\bibfield
  {title} {\bibinfo {title} {Entanglement phase transition induced by the
  non-hermitian skin effect},\ }\href
  {https://doi.org/10.1103/PhysRevX.13.021007} {\bibfield  {journal} {\bibinfo
  {journal} {Phys. Rev. X}\ }\textbf {\bibinfo {volume} {13}},\ \bibinfo
  {pages} {021007} (\bibinfo {year} {2023})}\BibitemShut {NoStop}%
\bibitem [{\citenamefont {Liu}\ \emph {et~al.}(2021{\natexlab{a}})\citenamefont
  {Liu}, \citenamefont {Wang}, \citenamefont {Zheng},\ and\ \citenamefont
  {Chen}}]{Chen21prb}%
  \BibitemOpen
  \bibfield  {author} {\bibinfo {author} {\bibfnamefont {Y.}~\bibnamefont
  {Liu}}, \bibinfo {author} {\bibfnamefont {Y.}~\bibnamefont {Wang}}, \bibinfo
  {author} {\bibfnamefont {Z.}~\bibnamefont {Zheng}},\ and\ \bibinfo {author}
  {\bibfnamefont {S.}~\bibnamefont {Chen}},\ }\bibfield  {title} {\bibinfo
  {title} {Exact non-hermitian mobility edges in one-dimensional quasicrystal
  lattice with exponentially decaying hopping and its dual lattice},\ }\href
  {https://doi.org/10.1103/PhysRevB.103.134208} {\bibfield  {journal} {\bibinfo
   {journal} {Phys. Rev. B}\ }\textbf {\bibinfo {volume} {103}},\ \bibinfo
  {pages} {134208} (\bibinfo {year} {2021}{\natexlab{a}})}\BibitemShut
  {NoStop}%
\bibitem [{\citenamefont {Liu}\ \emph {et~al.}(2020)\citenamefont {Liu},
  \citenamefont {Jiang}, \citenamefont {Cao},\ and\ \citenamefont
  {Chen}}]{PhysRevB.101.174205}%
  \BibitemOpen
  \bibfield  {author} {\bibinfo {author} {\bibfnamefont {Y.}~\bibnamefont
  {Liu}}, \bibinfo {author} {\bibfnamefont {X.-P.}\ \bibnamefont {Jiang}},
  \bibinfo {author} {\bibfnamefont {J.}~\bibnamefont {Cao}},\ and\ \bibinfo
  {author} {\bibfnamefont {S.}~\bibnamefont {Chen}},\ }\bibfield  {title}
  {\bibinfo {title} {Non-hermitian mobility edges in one-dimensional
  quasicrystals with parity-time symmetry},\ }\href
  {https://doi.org/10.1103/PhysRevB.101.174205} {\bibfield  {journal} {\bibinfo
   {journal} {Phys. Rev. B}\ }\textbf {\bibinfo {volume} {101}},\ \bibinfo
  {pages} {174205} (\bibinfo {year} {2020})}\BibitemShut {NoStop}%
\bibitem [{\citenamefont {Cai}(2021)}]{PhysRevB.103.214202}%
  \BibitemOpen
  \bibfield  {author} {\bibinfo {author} {\bibfnamefont {X.}~\bibnamefont
  {Cai}},\ }\bibfield  {title} {\bibinfo {title} {Localization and topological
  phase transitions in non-hermitian aubry-andr\'e-harper models with $p$-wave
  pairing},\ }\href {https://doi.org/10.1103/PhysRevB.103.214202} {\bibfield
  {journal} {\bibinfo  {journal} {Phys. Rev. B}\ }\textbf {\bibinfo {volume}
  {103}},\ \bibinfo {pages} {214202} (\bibinfo {year} {2021})}\BibitemShut
  {NoStop}%
\bibitem [{\citenamefont {Padhan}\ \emph {et~al.}(2024)\citenamefont {Padhan},
  \citenamefont {Padhi},\ and\ \citenamefont {Mishra}}]{Mishra}%
  \BibitemOpen
  \bibfield  {author} {\bibinfo {author} {\bibfnamefont {A.}~\bibnamefont
  {Padhan}}, \bibinfo {author} {\bibfnamefont {S.~R.}\ \bibnamefont {Padhi}},\
  and\ \bibinfo {author} {\bibfnamefont {T.}~\bibnamefont {Mishra}},\
  }\bibfield  {title} {\bibinfo {title} {Complete delocalization and reentrant
  topological transition in a non-hermitian quasiperiodic lattice},\ }\href
  {https://doi.org/10.1103/PhysRevB.109.L020203} {\bibfield  {journal}
  {\bibinfo  {journal} {Phys. Rev. B}\ }\textbf {\bibinfo {volume} {109}},\
  \bibinfo {pages} {L020203} (\bibinfo {year} {2024})}\BibitemShut {NoStop}%
\bibitem [{\citenamefont {Longhi}(2019)}]{Longhi2019}%
  \BibitemOpen
  \bibfield  {author} {\bibinfo {author} {\bibfnamefont {S.}~\bibnamefont
  {Longhi}},\ }\bibfield  {title} {\bibinfo {title} {Topological phase
  transition in non-hermitian quasicrystals},\ }\href
  {https://doi.org/10.1103/PhysRevLett.122.237601} {\bibfield  {journal}
  {\bibinfo  {journal} {Phys. Rev. Lett.}\ }\textbf {\bibinfo {volume} {122}},\
  \bibinfo {pages} {237601} (\bibinfo {year} {2019})}\BibitemShut {NoStop}%
\bibitem [{\citenamefont {Schiffer}\ \emph {et~al.}(2021)\citenamefont
  {Schiffer}, \citenamefont {Liu}, \citenamefont {Hu},\ and\ \citenamefont
  {Wang}}]{HuHui}%
  \BibitemOpen
  \bibfield  {author} {\bibinfo {author} {\bibfnamefont {S.}~\bibnamefont
  {Schiffer}}, \bibinfo {author} {\bibfnamefont {X.-J.}\ \bibnamefont {Liu}},
  \bibinfo {author} {\bibfnamefont {H.}~\bibnamefont {Hu}},\ and\ \bibinfo
  {author} {\bibfnamefont {J.}~\bibnamefont {Wang}},\ }\bibfield  {title}
  {\bibinfo {title} {Anderson localization transition in a robust
  $\mathcal{PT}$-symmetric phase of a generalized aubry-andr\'e model},\ }\href
  {https://doi.org/10.1103/PhysRevA.103.L011302} {\bibfield  {journal}
  {\bibinfo  {journal} {Phys. Rev. A}\ }\textbf {\bibinfo {volume} {103}},\
  \bibinfo {pages} {L011302} (\bibinfo {year} {2021})}\BibitemShut {NoStop}%
\bibitem [{\citenamefont {Gandhi}\ and\ \citenamefont
  {Bandyopadhyay}(2023)}]{Gandhi}%
  \BibitemOpen
  \bibfield  {author} {\bibinfo {author} {\bibfnamefont {S.}~\bibnamefont
  {Gandhi}}\ and\ \bibinfo {author} {\bibfnamefont {J.~N.}\ \bibnamefont
  {Bandyopadhyay}},\ }\bibfield  {title} {\bibinfo {title} {Topological triple
  phase transition in non-hermitian quasicrystals with complex asymmetric
  hopping},\ }\href {https://doi.org/10.1103/PhysRevB.108.014204} {\bibfield
  {journal} {\bibinfo  {journal} {Phys. Rev. B}\ }\textbf {\bibinfo {volume}
  {108}},\ \bibinfo {pages} {014204} (\bibinfo {year} {2023})}\BibitemShut
  {NoStop}%
\bibitem [{\citenamefont {Lin}\ \emph {et~al.}(2022)\citenamefont {Lin},
  \citenamefont {Li}, \citenamefont {Xiao}, \citenamefont {Wang}, \citenamefont
  {Yi},\ and\ \citenamefont {Xue}}]{XueP}%
  \BibitemOpen
  \bibfield  {author} {\bibinfo {author} {\bibfnamefont {Q.}~\bibnamefont
  {Lin}}, \bibinfo {author} {\bibfnamefont {T.}~\bibnamefont {Li}}, \bibinfo
  {author} {\bibfnamefont {L.}~\bibnamefont {Xiao}}, \bibinfo {author}
  {\bibfnamefont {K.}~\bibnamefont {Wang}}, \bibinfo {author} {\bibfnamefont
  {W.}~\bibnamefont {Yi}},\ and\ \bibinfo {author} {\bibfnamefont
  {P.}~\bibnamefont {Xue}},\ }\bibfield  {title} {\bibinfo {title} {Topological
  phase transitions and mobility edges in non-hermitian quasicrystals},\ }\href
  {https://doi.org/10.1103/PhysRevLett.129.113601} {\bibfield  {journal}
  {\bibinfo  {journal} {Phys. Rev. Lett.}\ }\textbf {\bibinfo {volume} {129}},\
  \bibinfo {pages} {113601} (\bibinfo {year} {2022})}\BibitemShut {NoStop}%
\bibitem [{\citenamefont {Liu}\ \emph {et~al.}(2021{\natexlab{b}})\citenamefont
  {Liu}, \citenamefont {Wang}, \citenamefont {Liu}, \citenamefont {Zhou},\ and\
  \citenamefont {Chen}}]{LiuYX2021a}%
  \BibitemOpen
  \bibfield  {author} {\bibinfo {author} {\bibfnamefont {Y.}~\bibnamefont
  {Liu}}, \bibinfo {author} {\bibfnamefont {Y.}~\bibnamefont {Wang}}, \bibinfo
  {author} {\bibfnamefont {X.-J.}\ \bibnamefont {Liu}}, \bibinfo {author}
  {\bibfnamefont {Q.}~\bibnamefont {Zhou}},\ and\ \bibinfo {author}
  {\bibfnamefont {S.}~\bibnamefont {Chen}},\ }\bibfield  {title} {\bibinfo
  {title} {Exact mobility edges, $\mathcal{PT}$-symmetry breaking, and skin
  effect in one-dimensional non-hermitian quasicrystals},\ }\href
  {https://doi.org/10.1103/PhysRevB.103.014203} {\bibfield  {journal} {\bibinfo
   {journal} {Phys. Rev. B}\ }\textbf {\bibinfo {volume} {103}},\ \bibinfo
  {pages} {014203} (\bibinfo {year} {2021}{\natexlab{b}})}\BibitemShut
  {NoStop}%
\bibitem [{\citenamefont {Liu}\ \emph {et~al.}(2021{\natexlab{c}})\citenamefont
  {Liu}, \citenamefont {Zhou},\ and\ \citenamefont {Chen}}]{Liuyx2021}%
  \BibitemOpen
  \bibfield  {author} {\bibinfo {author} {\bibfnamefont {Y.}~\bibnamefont
  {Liu}}, \bibinfo {author} {\bibfnamefont {Q.}~\bibnamefont {Zhou}},\ and\
  \bibinfo {author} {\bibfnamefont {S.}~\bibnamefont {Chen}},\ }\bibfield
  {title} {\bibinfo {title} {Localization transition, spectrum structure, and
  winding numbers for one-dimensional non-hermitian quasicrystals},\ }\href
  {https://doi.org/10.1103/PhysRevB.104.024201} {\bibfield  {journal} {\bibinfo
   {journal} {Phys. Rev. B}\ }\textbf {\bibinfo {volume} {104}},\ \bibinfo
  {pages} {024201} (\bibinfo {year} {2021}{\natexlab{c}})}\BibitemShut
  {NoStop}%
\bibitem [{\citenamefont {Acharya}\ \emph {et~al.}(2022)\citenamefont
  {Acharya}, \citenamefont {Chakrabarty}, \citenamefont {Sahu},\ and\
  \citenamefont {Datta}}]{Datta}%
  \BibitemOpen
  \bibfield  {author} {\bibinfo {author} {\bibfnamefont {A.~P.}\ \bibnamefont
  {Acharya}}, \bibinfo {author} {\bibfnamefont {A.}~\bibnamefont
  {Chakrabarty}}, \bibinfo {author} {\bibfnamefont {D.~K.}\ \bibnamefont
  {Sahu}},\ and\ \bibinfo {author} {\bibfnamefont {S.}~\bibnamefont {Datta}},\
  }\bibfield  {title} {\bibinfo {title} {Localization, $\mathcal{PT}$ symmetry
  breaking, and topological transitions in non-hermitian quasicrystals},\
  }\href {https://doi.org/10.1103/PhysRevB.105.014202} {\bibfield  {journal}
  {\bibinfo  {journal} {Phys. Rev. B}\ }\textbf {\bibinfo {volume} {105}},\
  \bibinfo {pages} {014202} (\bibinfo {year} {2022})}\BibitemShut {NoStop}%
\bibitem [{\citenamefont {Zhou}(2023)}]{ZhouLW}%
  \BibitemOpen
  \bibfield  {author} {\bibinfo {author} {\bibfnamefont {L.}~\bibnamefont
  {Zhou}},\ }\bibfield  {title} {\bibinfo {title} {Non-abelian generalization
  of non-hermitian quasicrystals: $\mathcal{PT}$-symmetry breaking,
  localization, entanglement, and topological transitions},\ }\href
  {https://doi.org/10.1103/PhysRevB.108.014202} {\bibfield  {journal} {\bibinfo
   {journal} {Phys. Rev. B}\ }\textbf {\bibinfo {volume} {108}},\ \bibinfo
  {pages} {014202} (\bibinfo {year} {2023})}\BibitemShut {NoStop}%
\bibitem [{\citenamefont {Xia}\ \emph {et~al.}(2022)\citenamefont {Xia},
  \citenamefont {Huang}, \citenamefont {Wang},\ and\ \citenamefont
  {Li}}]{XiaX2022}%
  \BibitemOpen
  \bibfield  {author} {\bibinfo {author} {\bibfnamefont {X.}~\bibnamefont
  {Xia}}, \bibinfo {author} {\bibfnamefont {K.}~\bibnamefont {Huang}}, \bibinfo
  {author} {\bibfnamefont {S.}~\bibnamefont {Wang}},\ and\ \bibinfo {author}
  {\bibfnamefont {X.}~\bibnamefont {Li}},\ }\bibfield  {title} {\bibinfo
  {title} {Exact mobility edges in the non-hermitian
  ${t}_{1}\text{\ensuremath{-}}{t}_{2}$ model: Theory and possible experimental
  realizations},\ }\href {https://doi.org/10.1103/PhysRevB.105.014207}
  {\bibfield  {journal} {\bibinfo  {journal} {Phys. Rev. B}\ }\textbf {\bibinfo
  {volume} {105}},\ \bibinfo {pages} {014207} (\bibinfo {year}
  {2022})}\BibitemShut {NoStop}%
\bibitem [{\citenamefont {Chen}\ \emph {et~al.}(2022)\citenamefont {Chen},
  \citenamefont {Cheng}, \citenamefont {Lin}, \citenamefont {Asgari},\ and\
  \citenamefont {Xianlong}}]{PhysRevB.106.144208}%
  \BibitemOpen
  \bibfield  {author} {\bibinfo {author} {\bibfnamefont {W.}~\bibnamefont
  {Chen}}, \bibinfo {author} {\bibfnamefont {S.}~\bibnamefont {Cheng}},
  \bibinfo {author} {\bibfnamefont {J.}~\bibnamefont {Lin}}, \bibinfo {author}
  {\bibfnamefont {R.}~\bibnamefont {Asgari}},\ and\ \bibinfo {author}
  {\bibfnamefont {G.}~\bibnamefont {Xianlong}},\ }\bibfield  {title} {\bibinfo
  {title} {Breakdown of the correspondence between the real-complex and
  delocalization-localization transitions in non-hermitian quasicrystals},\
  }\href {https://doi.org/10.1103/PhysRevB.106.144208} {\bibfield  {journal}
  {\bibinfo  {journal} {Phys. Rev. B}\ }\textbf {\bibinfo {volume} {106}},\
  \bibinfo {pages} {144208} (\bibinfo {year} {2022})}\BibitemShut {NoStop}%
\bibitem [{\citenamefont {Avila}(2015)}]{avila}%
  \BibitemOpen
  \bibfield  {author} {\bibinfo {author} {\bibfnamefont {A.}~\bibnamefont
  {Avila}},\ }\bibfield  {title} {\bibinfo {title} {{Global theory of
  one-frequency Schr\"{o}dinger operators}},\ }\href
  {https://doi.org/10.1007/s11511-015-0128-7} {\bibfield  {journal} {\bibinfo
  {journal} {Acta Mathematica}\ }\textbf {\bibinfo {volume} {215}},\ \bibinfo
  {pages} {1 } (\bibinfo {year} {2015})}\BibitemShut {NoStop}%
\bibitem [{\citenamefont {Wang}\ \emph
  {et~al.}(2023{\natexlab{b}})\citenamefont {Wang}, \citenamefont {Xia},
  \citenamefont {You}, \citenamefont {Zheng},\ and\ \citenamefont
  {Zhou}}]{zhouqiwang2023}%
  \BibitemOpen
  \bibfield  {author} {\bibinfo {author} {\bibfnamefont {Y.}~\bibnamefont
  {Wang}}, \bibinfo {author} {\bibfnamefont {X.}~\bibnamefont {Xia}}, \bibinfo
  {author} {\bibfnamefont {J.}~\bibnamefont {You}}, \bibinfo {author}
  {\bibfnamefont {Z.}~\bibnamefont {Zheng}},\ and\ \bibinfo {author}
  {\bibfnamefont {Q.}~\bibnamefont {Zhou}},\ }\bibfield  {title} {\bibinfo
  {title} {Exact mobility edges for 1d quasiperiodic models},\ }\href
  {https://doi.org/10.1007/s00220-023-04695-9} {\bibfield  {journal} {\bibinfo
  {journal} {Communications in Mathematical Physics}\ }\textbf {\bibinfo
  {volume} {401}},\ \bibinfo {pages} {2521} (\bibinfo {year}
  {2023}{\natexlab{b}})}\BibitemShut {NoStop}%
\bibitem [{\citenamefont {Hiramoto}\ and\ \citenamefont
  {Kohmoto}(1989)}]{Hiramoto}%
  \BibitemOpen
  \bibfield  {author} {\bibinfo {author} {\bibfnamefont {H.}~\bibnamefont
  {Hiramoto}}\ and\ \bibinfo {author} {\bibfnamefont {M.}~\bibnamefont
  {Kohmoto}},\ }\bibfield  {title} {\bibinfo {title} {Scaling analysis of
  quasiperiodic systems: Generalized harper model},\ }\href
  {https://doi.org/10.1103/PhysRevB.40.8225} {\bibfield  {journal} {\bibinfo
  {journal} {Phys. Rev. B}\ }\textbf {\bibinfo {volume} {40}},\ \bibinfo
  {pages} {8225} (\bibinfo {year} {1989})}\BibitemShut {NoStop}%
\bibitem [{\citenamefont {Evers}\ and\ \citenamefont
  {Mirlin}(2000)}]{PhysRevLett.84.3690}%
  \BibitemOpen
  \bibfield  {author} {\bibinfo {author} {\bibfnamefont {F.}~\bibnamefont
  {Evers}}\ and\ \bibinfo {author} {\bibfnamefont {A.~D.}\ \bibnamefont
  {Mirlin}},\ }\bibfield  {title} {\bibinfo {title} {Fluctuations of the
  inverse participation ratio at the anderson transition},\ }\href
  {https://doi.org/10.1103/PhysRevLett.84.3690} {\bibfield  {journal} {\bibinfo
   {journal} {Phys. Rev. Lett.}\ }\textbf {\bibinfo {volume} {84}},\ \bibinfo
  {pages} {3690} (\bibinfo {year} {2000})}\BibitemShut {NoStop}%
\bibitem [{\citenamefont {Gon\ifmmode~\mbox{\c{c}}\else \c{c}\fi{}alves}\ \emph
  {et~al.}(2020)\citenamefont {Gon\ifmmode~\mbox{\c{c}}\else \c{c}\fi{}alves},
  \citenamefont {Ribeiro}, \citenamefont {Castro},\ and\ \citenamefont
  {Ara\'ujo}}]{PhysRevLett.124.136405}%
  \BibitemOpen
  \bibfield  {author} {\bibinfo {author} {\bibfnamefont {M.}~\bibnamefont
  {Gon\ifmmode~\mbox{\c{c}}\else \c{c}\fi{}alves}}, \bibinfo {author}
  {\bibfnamefont {P.}~\bibnamefont {Ribeiro}}, \bibinfo {author} {\bibfnamefont
  {E.~V.}\ \bibnamefont {Castro}},\ and\ \bibinfo {author} {\bibfnamefont
  {M.~A.~N.}\ \bibnamefont {Ara\'ujo}},\ }\bibfield  {title} {\bibinfo {title}
  {Disorder-driven multifractality transition in weyl nodal loops},\ }\href
  {https://doi.org/10.1103/PhysRevLett.124.136405} {\bibfield  {journal}
  {\bibinfo  {journal} {Phys. Rev. Lett.}\ }\textbf {\bibinfo {volume} {124}},\
  \bibinfo {pages} {136405} (\bibinfo {year} {2020})}\BibitemShut {NoStop}%
\bibitem [{\citenamefont {Wang}\ \emph {et~al.}(2016)\citenamefont {Wang},
  \citenamefont {Liu}, \citenamefont {Xianlong},\ and\ \citenamefont
  {Hu}}]{PhysRevB.93.104504}%
  \BibitemOpen
  \bibfield  {author} {\bibinfo {author} {\bibfnamefont {J.}~\bibnamefont
  {Wang}}, \bibinfo {author} {\bibfnamefont {X.-J.}\ \bibnamefont {Liu}},
  \bibinfo {author} {\bibfnamefont {G.}~\bibnamefont {Xianlong}},\ and\
  \bibinfo {author} {\bibfnamefont {H.}~\bibnamefont {Hu}},\ }\bibfield
  {title} {\bibinfo {title} {Phase diagram of a non-abelian
  aubry-andr\'e-harper model with $p$-wave superfluidity},\ }\href
  {https://doi.org/10.1103/PhysRevB.93.104504} {\bibfield  {journal} {\bibinfo
  {journal} {Phys. Rev. B}\ }\textbf {\bibinfo {volume} {93}},\ \bibinfo
  {pages} {104504} (\bibinfo {year} {2016})}\BibitemShut {NoStop}%
\bibitem [{\citenamefont {Lin}\ \emph {et~al.}(2023)\citenamefont {Lin},
  \citenamefont {Chen}, \citenamefont {Guo},\ and\ \citenamefont
  {Gong}}]{lin2023general}%
  \BibitemOpen
  \bibfield  {author} {\bibinfo {author} {\bibfnamefont {X.}~\bibnamefont
  {Lin}}, \bibinfo {author} {\bibfnamefont {X.}~\bibnamefont {Chen}}, \bibinfo
  {author} {\bibfnamefont {G.-C.}\ \bibnamefont {Guo}},\ and\ \bibinfo {author}
  {\bibfnamefont {M.}~\bibnamefont {Gong}},\ }\bibfield  {title} {\bibinfo
  {title} {General approach to the critical phase with coupled quasiperiodic
  chains},\ }\href {https://doi.org/10.1103/PhysRevB.108.174206} {\bibfield
  {journal} {\bibinfo  {journal} {Phys. Rev. B}\ }\textbf {\bibinfo {volume}
  {108}},\ \bibinfo {pages} {174206} (\bibinfo {year} {2023})}\BibitemShut
  {NoStop}%
\end{thebibliography}
\end{document}